\documentclass[%
 jmp,%
 unsortedaddress,
 amsmath,amssymb,
reprint,%
]{revtex4-1}

\usepackage[acronym,symbols,nogroupskip,nomain,nonumberlist,nopostdot,toc]{glossaries}
\usepackage{hyperref}
\usepackage[version=4]{mhchem}
\usepackage[english]{babel}
\usepackage{braket}
\usepackage{booktabs}
\usepackage{mathtools}
\usepackage{tikz}
\usepackage{graphicx}
\usepackage{dcolumn}
\usepackage{bm}

\newcommand{\pyscf}{\textsc{PySCF}}
\newcommand{\qiskit}{\textsc{Qiskit}}
\newcommand{\vmd}{\textsc{VMD}}

\newcommand{\hamil}{\hat{\mathcal{H}}}

\newcommand{\crop}[1]{\hat{a}^{\dagger}_{#1}}
\newcommand{\anop}[1]{\hat{a}_{#1}}

\makeatletter
\DeclareRobustCommand{\rvdots}{%
  \vbox{
    \baselineskip4\p@\lineskiplimit\z@
    \kern-\p@
    \hbox{.}\hbox{.}\hbox{.}
  }}
\makeatother

\newcolumntype{@}{>{\global\let\currentrowstyle\relax}}
\newcolumntype{^}{>{\currentrowstyle}}

\newacronym{ao}{AO}{Atomic Orbital}
\newacronym{as}{AS}{Active Space}
\newacronym{cas}{CAS}{Complete Active Space}
\newacronym{casci}{CASCI}{Complete Active Space Configuration Interaction}
\newacronym{casscf}{CASSCF}{Complete Active Space Self--Consistent Field}
\newacronym{cc}{CC}{Coupled Cluster}
\newacronym{ccsd}{CCSD}{Coupled Cluster Singles and Doubles}
\newacronym{dft}{DFT}{Density Functional Theory}
\newacronym{dmet}{DMET}{Density Matrix Embedding Theory}
\newacronym{dmft}{DMFT}{Dynamical Mean Field Theory}
\newacronym{ecp}{ECP}{Effective Core Potential}
\newacronym{fci}{FCI}{Full Configuration Interaction}
\newacronym{hf}{HF}{Hartree--Fock}
\newacronym{ks}{KS}{Kohn--Sham}
\newacronym{lda}{LDA}{Local Density Approximation}
\newacronym{lr}{LR}{Long-Range}
\newacronym{mo}{MO}{Molecular Orbital}
\newacronym{rhf}{RHF}{Restricted Hartree--Fock}
\newacronym{rks}{RKS}{Restricted Kohn--Sham}
\newacronym{rs}{RS}{Range--Separation}
\newacronym{scf}{SCF}{Self--Consistent Field}
\newacronym{si}{SI}{Supplementary Information}
\newacronym{sr}{SR}{Short-Range}
\newacronym{uccsd}{q-UCCSD}{quantum Unitary Coupled Cluster Singles and Doubles}
\newacronym{uhf}{UHF}{Unrestricted Hartree--Fock}
\newacronym{vqe}{VQE}{Variational Quantum Eigensolver}
\newacronym{vwn}{VWN}{Vosko, Wilk, and Nusair}
\newacronym{wft}{WFT}{Wave Function Theory}
\newacronym{xcf}{XCF}{Exchange--Correlation Functional}

\AtBeginDocument{
\heavyrulewidth=.08em
\lightrulewidth=.05em
\cmidrulewidth=.03em
\belowrulesep=.15ex
\belowbottomsep=0pt
\aboverulesep=.4ex
\abovetopsep=5pt
\cmidrulesep=\doublerulesep
\cmidrulekern=.5em
\defaultaddspace=.5em
}

\begin{document}

\title
{Quantum HF/DFT-Embedding Algorithms for Electronic Structure Calculations: Scaling up to Complex Molecular Systems}


\author{Max Rossmannek}
\affiliation{ 
IBM Quantum, IBM Research -- Zurich,
8803 R{\"u}schlikon, Switzerland
}
\author{Panagiotis Kl. Barkoutsos}%
\affiliation{ 
IBM Quantum, IBM Research -- Zurich,
8803 R{\"u}schlikon, Switzerland
}
\author{Pauline J. Ollitrault}%
\affiliation{ 
IBM Quantum, IBM Research -- Zurich,
8803 R{\"u}schlikon, Switzerland; Laboratory of Physical Chemistry, ETH Zurich, 8093 Z{\"u}rich, Switzerland
}
\author{Ivano Tavernelli}%
\email{ita@zurich.ibm.com}
\affiliation{ 
IBM Quantum, IBM Research -- Zurich,
8803 R{\"u}schlikon, Switzerland
}

\date{\today}

\begin{abstract}
In the near future, material and drug design may be aided by quantum computer assisted simulations.
These have the potential to target chemical systems intractable by the most powerful classical computers. 
However, the resources offered by contemporary quantum computers are still limited, restricting the chemical simulations to very simple molecules.
In order to rapidly scale up to more interesting molecular systems, we propose the embedding of the quantum  electronic structure calculation into a classically computed environment obtained at the \gls{hf} or \gls{dft} level of theory. 
We achieve this by constructing an effective Hamiltonian that incorporates a mean field potential describing the action of the inactive electrons on a selected \gls{as}.
The ground state of the AS Hamiltonian is determined by means of the \gls{vqe} algorithm.
With the proposed iterative \gls{dft} embedding scheme we are able to obtain energy correction terms for a single pyridine molecule that outperform the \gls{casscf} results regardless of the chosen \gls{as}.
\end{abstract}

\pacs{Valid PACS appear here}

\maketitle

\glsresetall

\section{Introduction}
Quantum chemistry simulations allow the prediction of important chemical processes throughout, for instance, the elucidation of reaction mechanisms by means of the calculation of ground or excited state electronic structure properties~\cite{Whitfield2011}.
A variety of research and industrial such as chemical catalysis, material design, drug discovery and photo-chemical applications for solar energy conversion, just to name a few~\cite{Helgaker2012, Vogiatzis2018}, could take advantage of these methods.
Since the development of the first computers, the research on quantum chemistry has blossomed and a large variety of algorithms has been developed aspiring to achieve more accurate solutions of Schr{\"o}dinger's equation.
However, despite many theoretical and algorithmic advances the solutions of many interesting and relevant problems in chemistry and physics remain out of reach due to the inherent exponential scaling of the Hilbert space associated with the electronic structure calculations.
While several approximate methods have been developed in the past to circumvent this issue, these often break down when considering strongly correlated systems such as transition metal complexes~\cite{Reiher2017} and complicated catalytic processes~\cite{Boero1998}.
In the past decades, quantum computing has emerged as a new potential computational paradigm for the solution of many problems in chemistry and physics for which classical algorithms have an unfavorable scaling.
In particular, quantum computing has been shown to be a useful resource in a variety of research areas such as chemistry~\cite{AspuruGuzik2018, Reiher2017}, drug discovery~\cite{Cao2018}, strongly correlated systems~\cite{Reiner2019,Sokolov2020}, field theory~\cite{Kokail2019, Mathis2020}, material science~\cite{Babbush2018a} and many others.

Despite these recent advances and the possibility to execute calculations on quantum devices (e.g., Ref.~\cite{IBMQ}), the application of quantum algorithms is still in its infancy.
In fact, most of the research in chemistry relies on hybrid quantum-classical algorithms~\cite{Wecker2015}, which use highly optimized classical (number crunching) functionalities together with quantum algorithms for the representation and optimization of the system wavefunction. 
The most well known quantum chemistry algorithm that provides the means to leverage state-of-the-art quantum hardware is the \gls{vqe}~\cite{Peruzzo2014}.

For the representation of the many-body wavefunction in quantum circuits, some of the approaches derived in quantum chemistry can be mapped directly to quantum computing.
In particular, the \gls{hf} method has proven to pose a useful starting point for the mapping  of electronic structure problems in the qubit space using the so-called \emph{second quantization} formalism. 
Among the most commonly used post-\gls{hf} expansions of the many-electron wavefunction in quantum computing is the \gls{cc} Ansatz~\cite{Cizek1966,Kummel2002,Bartlett2007}, which allows for a systematic and controlled inclusion of higher order configurations starting from the uncorrelated \gls{hf} Slater determinant. 
Several quantum implementation of \gls{cc} have already been introduced in the literature~\cite{Peruzzo2014, Omalley2016, Romero2018, Barkoutsos2018, Moll2018, Kuhn2019, Bauman2019, Evangelista2019}, including schemes for the optimization of the one-electron molecular basis functions~\cite{Sokolov2020, Mizukami2019}.
In addition to the classically-inspired expansions, pure native quantum representations of the many-electron wavefunctions that can be better optimized for the available quantum hardware platforms have been proposed~\cite{Kandala2017,Barkoutsos2018}. 
The combination of the \gls{vqe} algorithm with the different wavefunction Ans{\"a}tze showed already  interesting results in the calculation of ground~\cite{Omalley2016, Kandala2017, Hempel2018, Sagastizabal2019, Arute2020} and excited state properties~\cite{Ollitrault2019, Higgott2019, Parrish2019, nakanishi2019, ryabinkin2018, Mcclean2017, stair2019, santagati2018, jones2019, tilly2020, Peruzzo2014} of simple molecules (up to a few atoms).
However, this protocol does not allow to scale to larger systems using the currently available classical simulators of quantum circuits (limited to a maximum of about 50 qubits) or the available quantum computers (also limited to a few tens of qubits).
Therefore, in order to leverage the potential advantage of the available quantum algorithm we explore the possibility of an embedding scheme in which only a portion of the full system is represented by the high-level quantum computing approach, while the rest is treated with an efficient but (necessarily) approximated classical representation of the electronic structure, such as \gls{hf} or \gls{dft}.
This embedding approach is of particular relevance when the complex, highly correlated, subsystem can be localized in a well defined subspace of the complete set of one--electron orbitals used to represent the many-electron wavefunction.
In this case, an accurate description of the electronic structure is obtained at lower cost, namely $\mathcal{O}(N_{qc}^4)$ for the quantum computing ($qc$) subsystem and $\mathcal{O}(N_{env}^2)$ to $\mathcal{O}(N_{env}^3)$ for the \textit{environment} (env) (with $N_{env} = N_{tot} - N_{qc}$), instead of the $\mathcal{O}(N_{tot}^4)$ scaling when no embedding is used.

In this work, we propose \gls{hf} and \gls{dft}-based quantum embedding schemes based on the well known notion of an \gls{as}~\cite{Roos1980,Helgaker2000}, which defines the set of \textit{active} orbitals described by the quantum algorithm.
To this end, we will construct an effective Hamiltonian which incorporates a mean field potential of the \textit{inactive} electrons and, thus, fully replaces the explicit mapping of the corresponding orbitals in the quantum register.
The quantum algorithm is therefore restricted to a subset of active orbitals, which, however, feel the presence of the environment through the action of the mean field potential generated by the inactive electrons of the environment.
Similar approaches of the \gls{hf} embedding have been proposed in the literature~\cite{Bauer2016,Rubin2016} mainly based on \gls{dmft}~\cite{Georges1992} and \gls{dmet}~\cite{Knizia2012} for the high-level description of the subsystem.
The latter aims at a similar \gls{hf} embedding scheme. 
However, while the focus of its authors was the development of a self-consistent \gls{hf} embedding approach~\cite{Wu2020}, in this work we will only consider iterative embedding within the framework of DFT. 
Concerning the \gls{dmft} approach, this is based on Green's function techniques and therefore it is not particularly suited for the kind of molecular applications of interest to this work.  
Additionally, during the preparation of this paper another related approach appeared in the literature~\cite{Ma2020b}.
In this case, the authors propose a \gls{dft} embedding scheme similar to ours, which however uses a different Ansatz to resolve the double counting problem of the correlation terms.
Furthermore, they do not update the embedding potential in a self-consistent manner as we do in our work.

This paper is organized as follows.
In Section~\ref{sec:theory}, we outline the theory and the implementation of the proposed \gls{as} schemes for quantum electronic structure calculations embedded in HF and DFT.
We split the derivation into two parts: one for the \gls{hf} embedding scheme, and one for the \gls{dft} embedding scheme.
Section~\ref{sec:methods} lists the technical details of our numerical methods.
In Section~\ref{sec:results} we present and discuss results for both embedding schemes applied to a few molecular test systems, namely \ce{H2O}, \ce{N2}, \ce{O2}, \ce{CH2} and pyridine.
Section~\ref{sec:conclusions} summarizes and concludes.

\section{Theory}
\label{sec:theory}

In this work, we propose two embedding schemes for quantum electronic structure algorithms based on \gls{hf} and \gls{ks} \gls{dft} \glspl{mo}.
The subsystem solved by means of the quantum approach (such as \gls{uccsd}~\cite{Whitfield2011,Peruzzo2014,Barkoutsos2018}) is embedded in the potential generated by the environment (i.e., the remaining electrons), which is computed within the \gls{hf} or \gls{dft} framework. 
Our solutions are based on the \gls{rs} technique for the two--electron integrals~\cite{Savin1995}, which allows for a rigorous partitioning of the problem into a subsystem (i.e., the \gls{as}) and its environment.
If this partitioning is done wisely, we can achieve a good level of accuracy for many properties of interest while significantly reducing the costs of the calculation. 
Furthermore, in the case of the \gls{dft} embedding scheme (which is the main target of this work) we will extend the algorithm to include the self-consistent optimization of the embedding potential, leading to more accurate energies and densities.
In the following we will call \textit{active} electrons the electrons that are part of the \gls{as}, while the remaining ones will be referred to as \textit{inactive}.

\subsection{Hartree--Fock Embedding}
\label{sec:HF_embedding}

In this first section, we derive the so-called \emph{inactive Fock operator}.
The goal of this operator is to embed the quantum computation into a classically computed environment treated at the \gls{hf} level of theory, through the notion of an \gls{as}.
While this method is not new and different variants of it have been implemented before in other software packages~\cite{Takeshita2020,Urbanek2020,Ma2020b}, in the following we summarize the key concepts that are needed for its implementation within the framework of quantum computing in \qiskit{}~\cite{Qiskit}.
This section also lays down the fundamentals for the implementation of the \gls{dft} embedding scheme presented in Section~\ref{sec:DFT_embedding}.

\begin{figure}
    \centering
    \includegraphics{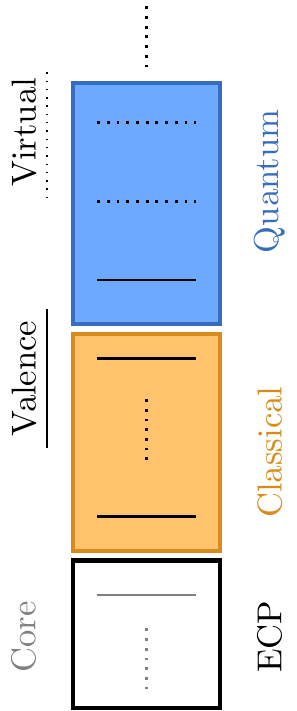}
    \caption{%
    Separation of the \glspl{mo} into active and inactive components.
    The active orbitals (blue box) are mapped onto the qubit space and treated with the \gls{uccsd} approach while the inactive ones (orange box) are part of the HF/DFT embedding and are evaluated classically.
    \glsfirstplural{ecp} can be used in replacement of all inactive core electrons (white box) with the aim of reducing the computational cost.
    }
    \label{fig:cas_comparison}
\end{figure}

The benefit of this embedding scheme lies in outsourcing the calculation of the inactive electrons to the classical \gls{hf} driver while the quantum computation is restricted to the \gls{as}.
In this way, less qubit resources are necessary to investigate the electronic energy of a molecular system, making the entire calculation much more efficient while keeping a good level of accuracy.
Fig.~\ref{fig:cas_comparison} depicts the separation of the orbitals into the active and inactive spaces.

The total electronic energy, $E$, is defined by expectation value of the system Hamiltonian, $\hamil$, 
\begin{align}
    E &= \Braket{\Psi|\hamil|\Psi}
    = \sum_{pq} h_{pq}D_{pq} + \frac{1}{2}\sum_{pqrs} g_{pqrs}d_{pqrs} \, ,
    \label{eq:schroedinger_2nd_quant}
\end{align}
where $\Psi$ is the wavefunction, $h_{pq}$ and $g_{pqrs}$ are the one-- and two--electron integrals, respectively, and $D$ and $d$ are the one-- and two--particle density matrices.

To achieve the implementation of the \gls{hf} embedding we split the one--electron density into an active and inactive part, $D=D^{A}+D^{I}$.
In the \gls{mo} basis, the latter simplifies  to $D^{I}_{iq}=2\delta_{iq}$, where we use Helgaker's notation of indices~\cite{Helgaker2000} in which $i,j,k,l$ denote \emph{inactive}, $u,v,x,y$ denote \emph{active} and $p,q,r,s$ denote \emph{general} \glspl{mo}.
As shown in the Appendix~\ref{app:rhf_embed_deriv} inserting this into Eq.~\eqref{eq:schroedinger_2nd_quant} leads to
\begin{align}
    E &= E^{I} + \sum_{uv} F^{I}_{uv} D_{uv}^{A} + \frac{1}{2} \sum_{uvxy} g_{uvxy}d_{uvxy}^{A} \, ,
    \label{eq:as_embedding}
\end{align}
where we define the \emph{inactive Fock operator},
\begin{align}
    F^{I}_{pq} &= h_{pq} + \sum_{i} \left(2g_{iipq} - g_{iqpi}\right), \label{eq:fock_inactive} \\
    \intertext{and the \emph{inactive energy},}
    E^{I} &= \sum_{j} h_{jj} + F^{I}_{jj} = \frac{1}{2} \sum_{ij} \left(h_{ij} + F^{I}_{ij}\right)D^{I}_{ij} \, . \label{eq:energy_inactive}
\end{align}
Comparing Eq.~\eqref{eq:schroedinger_2nd_quant} and Eq.~\eqref{eq:as_embedding} we observe the following differences.
In Eq.~\eqref{eq:as_embedding} the \emph{inactive} Fock operator, $F^{I}$, replaces the one-electron integrals, $h_{pq}$, the \emph{active} one-- and two--electron density matrices, $D^{A}$ and $d^{A}$, replace $D$ and $d$, and the constant energy offset, $E^{I}$, is added.

Therefore, the Hamiltonian which we will simulate on the quantum computer ($qc$) takes the form
\begin{align}
    \hamil_{qc} &= \sum_{uv} F^{I}_{uv} \crop{u}\anop{v} + \sum_{uvxy} g_{uvxy} \crop{u}\crop{v}\anop{x}\anop{y} \, ,
\end{align}
where $\crop{u}$ and $\anop{u}$ are the creation and annihilation Fermionic operators (later mapped to the qubit space using the parity transformation~\cite{Bravyi2002}).
Note that all indices are restricted to the \gls{as}, significantly reducing the required quantum resources. 

The extension for the unrestricted formalism is obtained in a similar manner and is outlined in Appendix~\ref{app:uhf_embed_deriv}.

\subsection{Density Functional Theory Embedding} \label{sec:DFT_embedding}

In order to extend the embedding to work with \gls{dft} we need to introduce a \gls{rs} of the two--electron integrals, $g_{pqrs}$~\cite{Savin1995}.
To this extent, we split the two--electron operator, $\hat{g}_{pq}$, into a \gls{lr} and a \gls{sr} part,
\begin{align}
    \hat{g}_{pq} &= \frac{1}{|\hat{r}_{p}-\hat{r}_{q}|}
    = \hat{g}^{\mu,\text{LR}}_{pq} + \hat{g}^{\mu,\text{SR}}_{pq},
    \label{eq:gpq_rs}
\end{align}
where $\mu$ is the \gls{rs} parameter of unit $a.u.^{-1}$.
This is necessary in order to avoid a double counting of the correlation terms which are present in both, \gls{dft} and \gls{wft}.
Since \gls{dft} is known to be accurate for \gls{sr} interactions~\cite{Savin1995}, we can use it to treat the \gls{sr} part while the \gls{lr} interactions are calculated with \gls{wft}.

Our derivation of the following equations follows that of Hedeg{\aa}rd et al.~\cite{Hedegard2015} closely.
Additionally, we provide our detailed derivations in Appendix~\ref{app:dft_embed_deriv}. \\
With the \gls{rs} of the two--electron integrals in place, we can split the total electronic energy into two terms,
\begin{align}
    E &= E^{\mu,\text{LR}}_{\text{WFT}} + E^{\mu,\text{SR}}_{\text{coul}+\text{xc},\text{DFT}} \, .
    \label{eq:adiab_conn}
\end{align}
Note that Eq.~\eqref{eq:adiab_conn} provides an \emph{adiabatic connection} between the pure \gls{dft} and the pure \gls{wft} solutions through the coupling parameter, $\mu$.
However, in order to simplify the notation we drop the superscript $\mu$ since it is anyways implied by the separation into \gls{lr} and \gls{sr}.

Analogous to Section \ref{sec:HF_embedding}, we can introduce an \gls{as} in the \gls{wft} part,
\begin{align}
    E &= E^{I,\text{LR}}_{\text{WFT}} + E^{A,\text{LR}}_{\text{WFT}} + E^{\text{SR}}_{\text{coul}+\text{xc},\text{DFT}} \, .
    \label{eq:as_adiab_conn}
\end{align}
Note that the difference between Eq.~\eqref{eq:as_embedding} and Eq.~\eqref{eq:as_adiab_conn} is that \gls{wft} only treats the \gls{lr} part.
Thus, the \emph{inactive Fock operator}, defined in Eq.~\eqref{eq:fock_inactive}, becomes
\begin{align}
    F^{I,\text{LR}}_{pq} &= h_{pq} + \sum_{i} \left(2g^{\text{LR}}_{iipq} - g^{\text{LR}}_{iqpi}\right).
    \label{eq:fock_inactive_lr}
\end{align}

In order to properly combine the \gls{sr}-\gls{dft} and \gls{lr}-\gls{wft} calculations we need to handle the non-linearity of $E^{\text{SR}}_{\text{coul}+\text{xc},\text{DFT}} = E^{\text{SR}}_{\text{coul}+\text{xc}}$ on the electronic density, $\rho$,
\begin{align}
    E_{\text{coul}+\text{xc}}^{\text{SR}}\left[\rho + \Delta\rho\right] &\neq
    E_{\text{coul}+\text{xc}}^{\text{SR}}\left[\rho\right] + E_{\text{coul}+\text{xc}}^{\text{SR}}\left[\Delta\rho\right],
    \label{eq:non_linearity}
\end{align}
where $\Delta\rho$ is the correction to the density obtained from the \gls{wft} calculation.
However, a linear model can be obtained with the following approximation,
\begin{align}
    E_{\text{coul}+\text{xc}}^{\text{SR}}&\left[\rho + \Delta\rho\right] - E_{\text{coul}+\text{xc}}^{\text{SR}}\left[\rho\right] \nonumber \\
    &\approx \int \frac{\delta E_{\text{coul}+\text{xc}}^{\text{SR}}}{\delta\rho(\vec{r})}\left[\rho\right]\Delta\rho(\vec{r})\text{d}\vec{r} \, .
    \label{eq:hxc_lin_mod}
\end{align}
The right hand side of Eq.~\eqref{eq:hxc_lin_mod} can then be expressed in terms of the \emph{Coulomb} integrals,
\begin{subequations}
    \begin{align}
        j^{\text{SR}}_{pq} &= \Braket{\phi_{p}|\frac{\delta E_{\text{coul}}^{\text{SR}}}{\delta\rho(\vec{r})}\left[\rho\right]|\phi_{q}}
        = \sum_{rs} g_{pqrs}^{\text{SR}} D_{rs} \\
        \shortintertext{and the \emph{exchange} integrals,}
        \nu^{\text{SR}}_{\text{xc},pq} &= \Braket{\phi_{p}|\frac{\delta E_{\text{xc}}^{\text{SR}}}{\delta\rho(\vec{r})}\left[\rho\right]|\phi_{q}}
        = \nu^{\text{SR}}_{\text{xc},pq}\left[\rho\right],
        \label{eq:vxc_sr}
    \end{align}
\end{subequations}
as,
\begin{align}
    \int \frac{\delta E_{\text{coul}+\text{xc}}^{\text{SR}}}{\delta\rho(\vec{r})}\left[\rho\right]\Delta\rho(\vec{r})\text{d}\vec{r}
    &= \sum_{pq} \left(j^{\text{SR}}_{pq} + \nu^{\text{SR}}_{\text{xc},pq}\right) \Delta D_{pq} \, .
    \label{eq:lin_mod_approximation}
\end{align}

Because of the non-linearity of Eq.~\eqref{eq:vxc_sr} the density needs to be updated in an iterative, self-consistent manner.
Therefore, we define the density and the density matrix at the iteration step $i$ as,
\begin{subequations}
    \begin{align}
        \rho^{(i+1)} &= \rho^{(i)} + \Delta\rho^{(i)}, \\
        D^{(i+1)}_{pq} &= D^{(i)}_{pq} + \Delta D_{pq}^{(i)} \, .
    \end{align}
    \label{eq:delta_rho}
\end{subequations}
This leads to the final form of the total electronic energy,
\begin{align}
    E &= \frac{1}{2}\sum_{ij} \left(
        h_{ij} + F^{I,\text{LR}}_{ij}
    \right) D^{I}_{ij}
    \nonumber \\
    &+ E_{\text{xc}}^{\text{SR}}\left[\rho^{(i)}\right]
    + \frac{1}{2} \sum_{ij} j^{I,\text{SR}}_{ij} D^{I}_{ij}
    \nonumber \\
    &- \sum_{uv} \left[\left(
        \frac{1}{2} j^{A,(i),\text{SR}}_{uv} + \nu^{\text{SR}}_{\text{xc},uv}\left[\rho^{(i)}\right]
    \right)D^{A,(i)}_{uv}\right. \nonumber \\
    &+ \left.\left(F^{I,\text{LR}}_{uv} + j^{I,\text{SR}}_{uv}
        + j^{A,(i),\text{SR}}_{uv} + \nu^{\text{SR}}_{\text{xc},uv}\left[\rho^{(i)}\right]
    \right) D^{A,(i+1)}_{uv} \right]
    \nonumber \\
    &+ \frac{1}{2} \sum_{uvxy} g_{uvxy}^{\text{LR}}d_{uvxy}^{A,(i+1)},
    \label{eq:rs_as_embedding_final}
\end{align}
where we have ordered the terms such that the top line contains all contributions which remain constant for the duration of the whole iterative procedure, the second and third lines correspond to the \gls{sr}-\gls{dft}, and the remaining lines correspond to the \gls{lr}-\gls{wft} energy terms, respectively.

\begin{figure}
    \centering
    \includegraphics{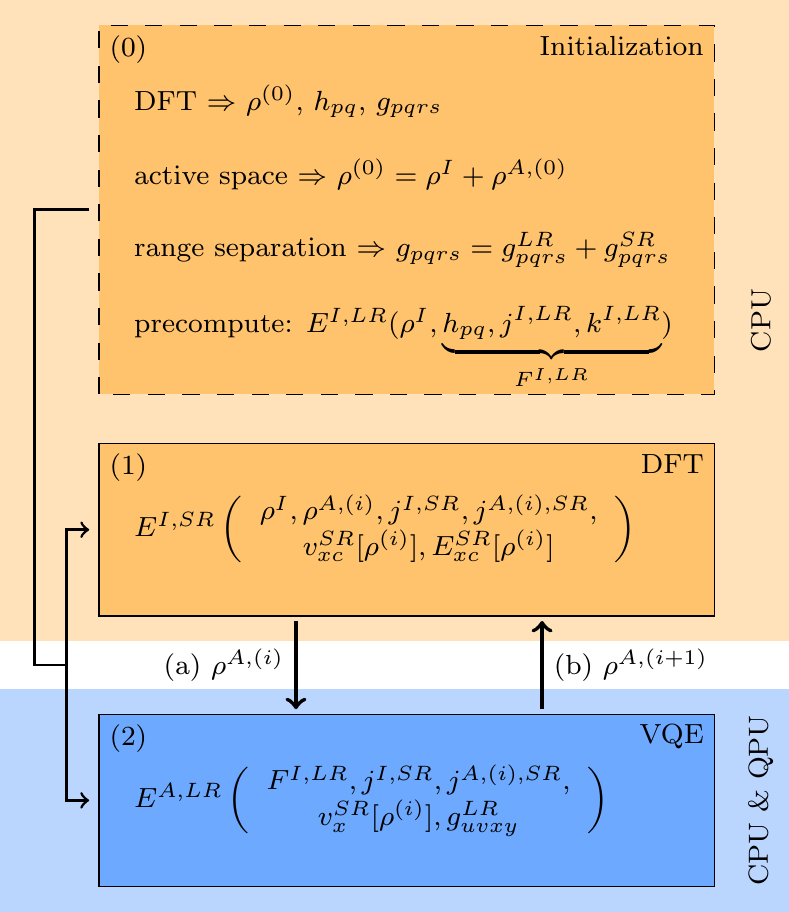}
    \caption{%
    Illustration of the \gls{dft} embedding scheme.
    During initialization $(0)$, a \gls{dft} calculation of the full system is performed using a classical code, providing the initial density, $\rho^{(0)}$, and the one-- and two-- electron integrals, $h_{pq}$ and $g_{pqrs}$.
    The density is then split into inactive, $I$, and active parts, $A$, and the two--electron integrals are separated into long-range, $LR$, and short-range, $SR$, components.
    In step $(1)$ the inactive \gls{sr} energy contribution is calculated at the \gls{dft} level of theory.
    The resulting `active' density component $(a)$ is used in step $(2)$ to initialize the \gls{vqe} optimization.
    This returns the active \gls{lr} energy contribution and the updated electronic density, which is used as a new input for the \gls{dft} calculation, $(b)$.
    Steps $(1)$ and $(2)$ are repeated until convergence.
    }
    \label{fig:iterative}
\end{figure}

Fig.~\ref{fig:iterative} summarizes the implementation of this \gls{dft} embedding scheme.
The initialization step includes all the pre-calculations and the computation of the constant \emph{inactive} \gls{lr} energy contribution (first line of Eq.~\eqref{eq:rs_as_embedding_final}).
The resulting energy terms of steps $(1)$ and $(2)$ in Fig.~\ref{fig:iterative} correspond to lines two and three, and four and five of Eq.~\eqref{eq:rs_as_embedding_final}, respectively.
These two calculations iterate, upon exchanging the active electronic density, $\rho^{A}$, until the total electronic energy reaches convergence.

\section{Numerical Methods}
\label{sec:methods}

The \gls{hf} and \gls{dft} embedding schemes have been implemented in the latest development version $0.8$ of \qiskit{} Aqua Chemistry.
The source code is made available in the Github repository~\cite{Qiskit_Aqua}.
For the classical computing backend we choose \pyscf{}~\cite{Sun2017} since it allows quick prototyping within Python, the same programming language used for \qiskit{}.

All the results presented hereafter are obtained by means of diagonalizing the Hamiltonian with the \texttt{ExactEigensolver}
\footnote{The \texttt{ExactEigensolver} method was renamed to \texttt{NumPyEigensolver} in \qiskit{} Aqua's $0.7$ release.}
algorithm as implemented in \qiskit{}.
In the case of the non-iterative \gls{hf} embedding scheme, we also run \gls{vqe} simulations with the \texttt{statevector} backend~\cite{Qiskit_Aqua}. 
This backend implements an exact, i.e. noiseless, simulation of the quantum circuit and, thus, is expected to converge to the same result as the \texttt{ExactEigensolver} approach when a suitable wavefunction Ansatz is chosen. 
In addition, for some selected systems (see below) we also perform  noisy \gls{vqe} calculations, referred to as \texttt{QASM} simulations, using noise models corresponding to the two IBM Q devices \textit{ibmq\_almaden} and \textit{ibmq\_boeblingen}.

\subsection{Hartree--Fock Embedding}
\label{sec:methods_HF_embedding}

In all simulations using the \gls{hf} embedding, we use the parity fermions-to-qubits mapping~\cite{Bravyi2002} and the \gls{uccsd} Ansatz~\cite{Barkoutsos2018} for the representation of the electronic wavefunction.
Furthermore, qubits are tapered off~\cite{Bravyi2017} in order to maximally reduce the computational costs.
The classical optimizers L-BFGS-B~\cite{Morales2011} and SLSQP~\cite{Kraft1988} are used for the optimization of the \gls{vqe} parameters in the case of noiseless and noisy simulations, respectively.

Ten qubits are needed for the simulation of the selected systems with the noisy \gls{vqe} algorithm, after application of the parity transformation. 
These could be reduced to six after tapering off~\cite{Bravyi2017}.
Thus, only a subset of the total qubits available on \textit{ibmq\_almaden} and \textit{ibmq\_boeblingen} was used to perform the simulations.
These subsets of qubits are \texttt{1,2,3,8,7,6} on \emph{ibmq\_almaden} and \texttt{3,2,1,6,7,8} on \emph{ibmq\_boeblingen}.
The connection between the first and last qubits of the sequence allows for a more efficient implementation of the quantum algorithm after compilation~\cite{Qiskit}.
This avoids the extensive use of \texttt{SWAP} gates to achieve coupling between qubits, which are not directly connected in the chip.

\subsection{Density Functional Theory Embedding}
\label{sec:methods_DFT_embedding}

In all DFT embedding applications we use the \gls{rs}-\glstext{xcf} (Range-Separated \glsdesc{xcf})\glsunset{xcf} \texttt{ldaerf} scheme~\cite{Toulouse2004a,Paziani2006} as implemented in the \texttt{xcfun} library~\cite{XCFun} for the separation of the \glstext{lda} (\glsdesc{lda})\glsunset{lda} functional into its short and long range components (see Section~\ref{sec:DFT_embedding}).
This approach achieves the splitting of the two--electron integrals by means of the error function which is a common approach in \gls{rs}-\gls{dft}~\cite{Savin1995,Heyd2003,Fromager2007}.
The use of the LDA functional is solely motivated by the current technical limitations of the \pyscf{} code.
Future extensions to allow the use of arbitrary \gls{dft} functionals are under investigation.
Nonetheless, the proposed scheme is fully independent from the nature of the selected functional and all applications presented in the following should be considered as proof-of-principle demonstrations extendable to any type of \gls{dft} functional.

In-line with the previous simulations with the \gls{hf} embedding scheme, we make use of the 6-31G* basis set, the parity mapping~\cite{Bravyi2002}, and the \gls{uccsd} Ansatz~\cite{Barkoutsos2018} for the representation of the electronic wavefunction.

\section{Results and Discussion}
\label{sec:results}

We present the results obtained with the proposed \gls{hf} and \gls{dft} embedding schemes (Section \ref{sec:HF_embedding} and \ref{sec:DFT_embedding}, respectively).
Using the \gls{hf} embedding approch, we investigate a variety of small molecules highlighting the broad applicability of the procedure. 
On the other hand, the better accuracy of \gls{dft} calculations over \gls{hf} will enable us to scale up 
to larger molecular systems such as the heterocyclic pyridine molecule. 

\subsection{Hartree--Fock Embedding}
\label{sec:results_HF_embed}

We test the \gls{hf} embedding scheme on several molecular systems, including \ce{H2O}, \ce{N2}, \ce{CH2} and \ce{O2}.
To simplify the discussion of the results, we take the water molecule as a benchmark system, while all other systems will be presented in the full paper. 

In the case of \ce{H2O}, we investigate the effect of the basis set on the accuracy of the ground state energy by increasing its size from STO-3G~\cite{Hehre1969} to 6-31G*~\cite{Hariharan1973} and cc-pVTZ~\cite{Dunning1989}. 
Furthermore, we consider several \glspl{as} ranging from the minimum of two electrons in two \glspl{mo} (CAS(2, 2)) all the way up to ten electrons in ten \glspl{mo} (CAS(10, 10)).

\subsubsection{A benchmark system: water}

\begin{figure}
    \centering
    \includegraphics{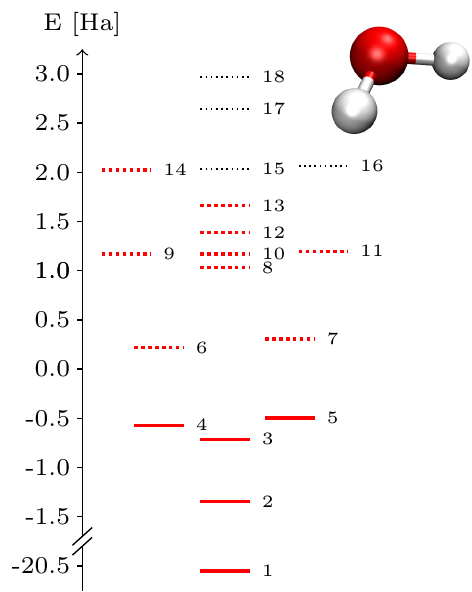}
    \caption{%
    Energy diagram of the \gls{hf}-\glspl{mo} of \ce{H2O} in the 6-31G* basis.
    Solid lines correspond to occupied \glspl{mo} while dotted lines represent virtual ones.
    The \glspl{mo} corresponding to the \emph{red} lines ($1$ to $14$) are visualized in Fig.~\ref{fig:orbitals_H2O_631Gst}.
    }
    \label{fig:MO_diagram_H2O_631Gst}
\end{figure}

\begin{figure}
    \centering
    \includegraphics[width= \columnwidth]{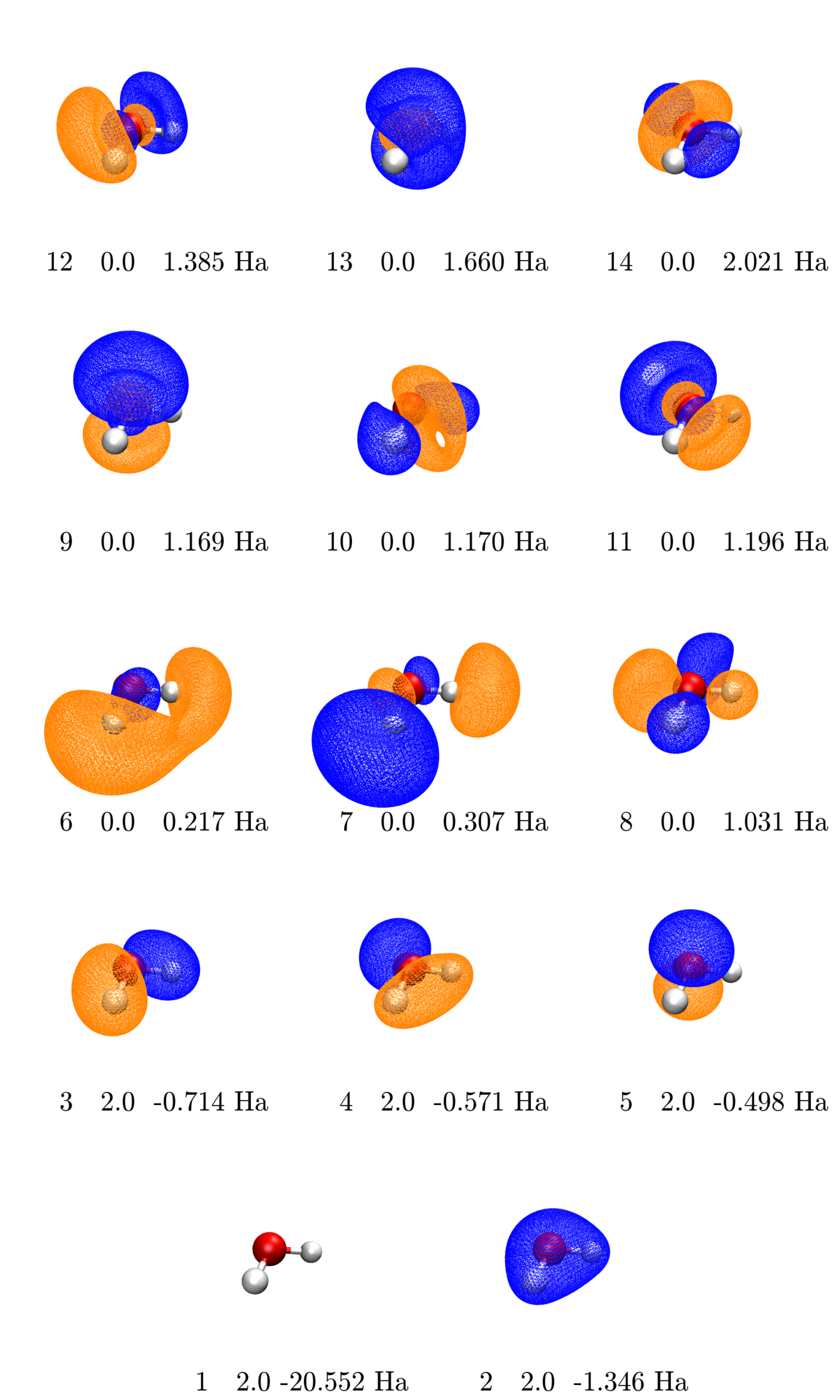}
    \caption{%
    Visualization of the active \gls{hf}-\glspl{mo} of water in the 6-31G* basis.
    The orbital coefficients, energies and occupation numbers are obtained with \pyscf{} using a \gls{rhf} calculation.
    All orbitals are rendered using \vmd{}~\cite{Humphrey1996} at an isovalue of $0.075$.
    The three numbers below each orbital correspond to the \gls{hf}-\glspl{mo} index (as indicated in Fig.~\ref{fig:MO_diagram_H2O_631Gst}), the occupation number, and the orbital energy.
    The first \gls{mo} is essentially identical to the oxygen's 1s orbital and completely hidden by the red sphere representing the oxygen atom.
    }
    \label{fig:orbitals_H2O_631Gst}
\end{figure}

As a first test case, we apply the \gls{hf} embedding scheme to the case of a single water molecule using the 6-31G* basis.
Fig.~\ref{fig:MO_diagram_H2O_631Gst} presents an energy diagram of the \gls{hf}-\gls{mo} energies and the active \gls{hf}-\glspl{mo} are shown in Fig.~\ref{fig:orbitals_H2O_631Gst}.

\begin{figure}
    \centering
    \includegraphics[width= \columnwidth]{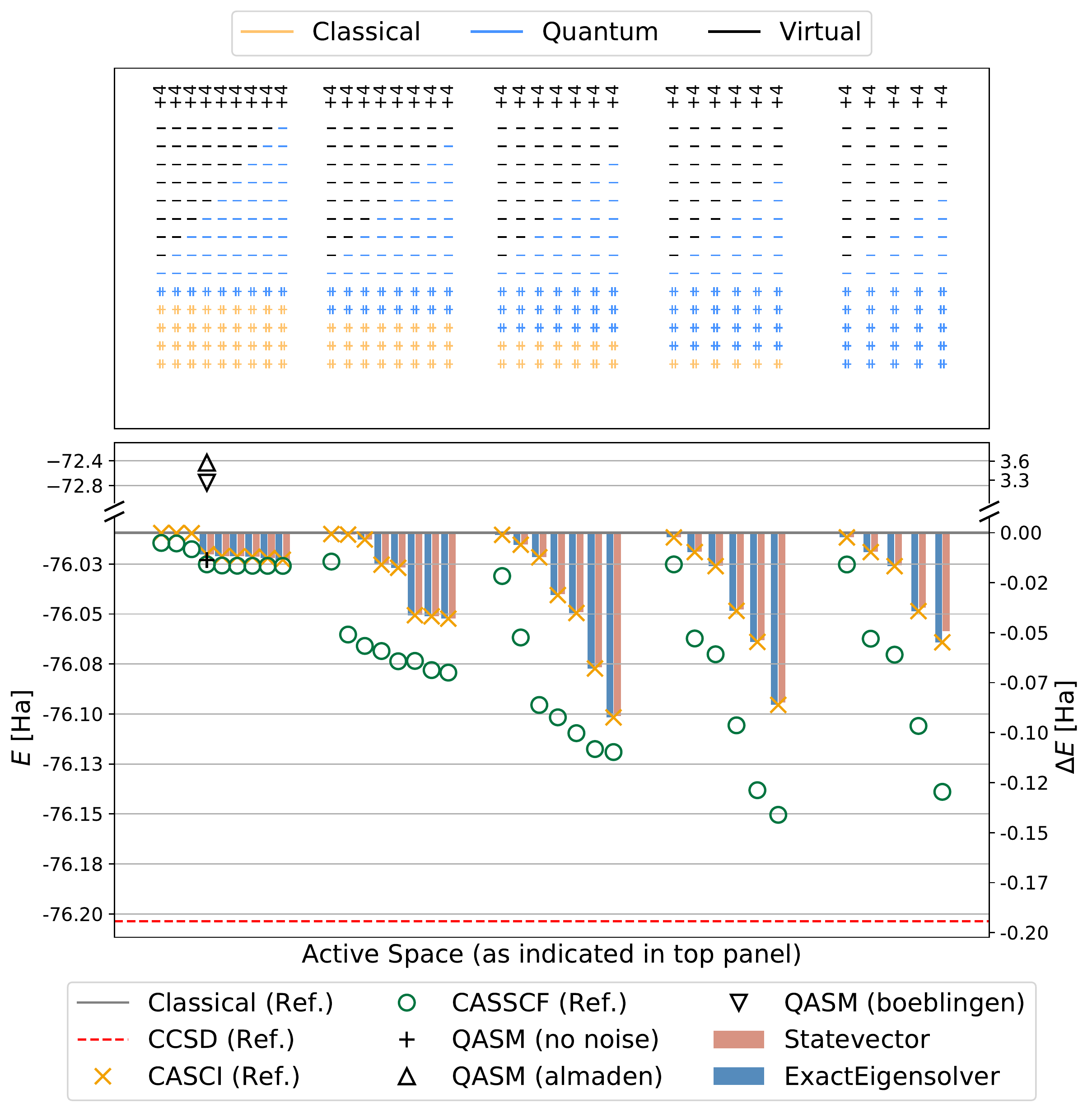}
    \caption{%
    Electronic structure energies (in Hartree) of a \textbf{\ce{H2O} molecule} obtained using the \gls{hf} embedding scheme for different choices of the AS.
    The classical HF reference was obtained with \gls{rhf} in the 6-31G* basis using \pyscf{}.
    All energy differences relative to this value are shown on the $y$-axis on the right (lower panel).
    The \gls{ccsd}, \gls{casci} and \gls{casscf} references were also computed with \pyscf{}.
    The corresponding \glspl{as} are depicted in the upper panel.
    The $+4$ above the \glspl{as} indicates that $4$ additional virtual orbitals are omitted from the visualization.
    The coloring follows the same scheme used in Figs.~\ref{fig:cas_comparison} and~\ref{fig:iterative}: orange for the inactive DFT orbitals that defines the embedding, blue the orbitals belonging to the AS, and in black the remaining virtual ones.
    }
    \label{fig:basic_H2O_RHF_631Gst}
\end{figure}

In Fig.~\ref{fig:basic_H2O_RHF_631Gst} we summarize the main results. 
In general, the \gls{uccsd} Ansatz applied in this work reproduces ground state energies in qualitative good agreement with the classical \gls{casci} approach~\cite{Sun2017} (orange crosses in Fig.~\ref{fig:basic_H2O_RHF_631Gst}).
This shows that the inclusion of merely the single and double excitations is sufficient to obtain the \gls{fci} accuracy within a given \gls{as} for a simple system such as \ce{H2O}.
Furthermore, most \texttt{statevector}-based \gls{vqe} calculations (brown bars in Fig.~\ref{fig:basic_H2O_RHF_631Gst}) converge to the exact solutions (blue bars in Fig.~\ref{fig:basic_H2O_RHF_631Gst}).
The \gls{casscf}~\cite{Sun2017} results (green circles in Fig.~\ref{fig:basic_H2O_RHF_631Gst}) are consistently lower (or equal) in energy than the \gls{casci} approach since this method also optimizes the orbital coefficients, which are kept fixed in the \gls{casci} method.

As expected~\cite{Veryazov2011,Stein2016}, we also observe that the energy corrections obtained with the embedding method do not trivially depend on the chosen \gls{as}, in particular with respect to the number of occupied versus unoccupied orbitals (see the trends in Fig.~\ref{fig:basic_H2O_RHF_631Gst}).
Thus, it is not possible to draw a general conclusion about the optimal selection of the \gls{as} since this is highly dependent on the molecule, the basis set, and the underlying optimization method.
In practice, other choices than the selection of the \glspl{mo} around the Fermi level could be also considered.
Indeed, tools have been developed which aim at automatizing the selection of the optimal \gls{as}~\cite{Stein2016,Stein2019,Faulstich2019,Sayfutyarova2017,Sagastizabal2019}.
However, the combination of these approaches with the proposed embedding scheme goes beyond the scope of this work and will become the subject of future investigations.

In order to assess the influence of the quantum sampling error on the accuracy of the embedding scheme, we perform some additional calculations using the \texttt{QASM}-based \gls{vqe} with a realistic description of the hardware noise. 
The results of these simulations for the CAS(2, 5) system are also included in  Fig.~\ref{fig:basic_H2O_RHF_631Gst} (black plus and triangles).
The main hardware noise sources are related to infidelity of the qubit operations, qubit decoherence, readout errors, and statistical sampling of the expectation values.
Concerning this last point, in our simulations we will use the standard value of $8192$ measurements for each expectation value. 
First, we start with a test \texttt{QASM}-based \gls{vqe} simulation in which we artificially repress all noise sources. 
For the selected AS, this `noiseless' simulation converges to the qualitatively correct energy value (black plus in Fig.~\ref{fig:basic_H2O_RHF_631Gst}), in agreement with the corresponding \texttt{statevector}-based \gls{vqe} calculation.
Upon addition of hardware noise models of the \emph{ibmq\_almaden} and \emph{ibmq\_boeblingen} devices the quality of the results drops dramatically, as indicated by the black triangles in Fig.~\ref{fig:basic_H2O_RHF_631Gst}.
Despite the careful selection of qubits on each device (cf.\@ Section~\ref{sec:methods_HF_embedding}), the circuit depth required for the implementation of the \gls{uccsd}-based embedding calculation is still beyond the limits of what can be executed (and simulated) on state-of-the-art quantum computers. 
This is reflected by the very large deviations between the \gls{vqe} energy and the actual ground state energy.
(In fact, the results are less accurate than the initial \gls{hf} energies.)

In general, these results on `noisy' simulations confirm the inadequacy of the \gls{uccsd} Ansatz for experiments on hardware, as reported for instance in Ref.~\cite{Ganzhorn2019}, and alternative wave function Ans{\"a}tze need to be investigated.
To this extent, a particle conserving heuristic Ansatz can become a viable option aiming at significantly shorter circuit depths~\cite{Kandala2017,Barkoutsos2018,Choquette2020}.

In the \gls{si} we also provide the outcomes obtained with the STO-3G and cc-pVTZ basis sets.
In the case of STO-3G the largest \gls{as} includes all available orbitals resulting in the recovery of the exact \gls{ccsd} energy (CAS(10, 7) in Fig.~S1).
Furthermore, we also perform another set of \texttt{QASM}-based \gls{vqe} simulations for the CAS(8, 5) case (black plus and triangles in Fig.~S1) which do not differ qualitatively from the results discussed previously for CAS(2, 5) in the 6-31G* basis.
In the case of the cc-pVTZ basis set, the quantitative energy correction terms are smaller than for 6-31G* due to the much larger total number of molecular orbitals (while keeping a constant AS size).

\subsubsection{Towards more complex systems}

\begin{table*}
    \centering
    \caption{%
    Summary of the \gls{hf}-based embedding calculations yielding the largest energy correction, $\varepsilon^{emb}_{AS}$ (cf.\@ Eq.~\eqref{eq:e_corr}).
    For systems with a non-singlet ground state we denote the number of active $\alpha$-electrons explicitly as CAS($N_{\text{el}}$ ($N_{\text{el}}^{\alpha}$), $N_{\text{mo}}$).
    All reference energies were obtained with the same initial \gls{hf} densities used for the embedding calculation and were computed with \pyscf{}.
    Energy values are in Ha.
    }
    \label{tab:basic_results_overview}
    \begin{tabular}{lllrrrrr}
        \toprule
        \addlinespace%
        System & Basis Set & Active Space &
        $E_{\text{q-UCCSD}}^{\text{HF}}$  ($\varepsilon_{\text{q-UCCSD}}^{\text{HF}}$) &
        $E_{\text{CASCI}}^{\text{HF}}$  &
        $E_{\text{CASSCF}}^{\text{HF}}$  ($\varepsilon_{\text{CASSCF}}^{\text{HF}}$) &
        $E_{\text{HF}}$  &
        $E_{\text{CCSD}}$ \\
        \addlinespace%
        \midrule
        \ce{H2O} & STO-3G & CAS(10, 7) & $-75.009$ ($100.0$) & $-75.009$ & $-75.009$ ($100.0$) & $-74.961$ & $-75.009$ \\
        \ce{H2O} & 6-31G* & CAS(6, 10) & $-76.102$ ($47.7$) & $-76.102$ & $-76.119$ ($56.4$) & $-76.009$ & $-76.204$ \\
        \ce{H2O} & cc-pVTZ & CAS(6, 10) & $-76.108$ ($17.9$) & $-76.108$ & $-76.191$ ($47.5$) & $-76.058$ & $-76.338$ \\
        \ce{N2} & 6-31G* & CAS(8, 10) & $-109.033$ ($29.2$) & $-109.033$ & $-109.114$ ($55.5$) & $-108.943$ & $-109.251$ \\
        \ce{CH2} & 6-31G* & CAS(6 (4), 10) & $-38.959$ ($37.6$) & $-38.959$ & $-38.992$ ($70.3$) & $-38.921$ & $-39.022$ \\
        \ce{O2} & 6-31G* & CAS(8 (5), 10) & $-149.707$ ($27.8$) & $-149.707$ & $-149.749$ ($40.7$) & $-149.616$ & $-149.943$ \\
        \bottomrule
    \end{tabular}
\end{table*}

In addition to the \ce{H2O} benchmark system, we apply the \gls{hf} embedding scheme to a series of other small molecules: \ce{N2}, \ce{CH2} and \ce{O2}, using the 6-31G* basis set.
These systems vary in size, spin multiplicity of the ground state and elemental composition and serve as a  demonstration of the wide applicability of our approach.
We summarize the main results of these calculations in Table~\ref{tab:basic_results_overview} where we also introduce the following measure for the energy correction
\begin{align}
    \varepsilon^{emb}_{AS} &= \frac{E^{emb}_{AS} - E_{\text{HF}}}{E_{\text{CCSD}} - E_{\text{HF}}} [\%] \, ,
    \label{eq:e_corr}
\end{align}
where the subscript, $AS$, and superscript, $emb$, indicate the method used to treat the \gls{as} and the `embedding` method, respectively.
Thus, $E_{\text{q-UCCSD}}^{\text{HF}}$ denotes the \gls{hf} embedding energies corresponding to Eq.~\eqref{eq:as_embedding} and $E_{\text{CASSCF}}^{\text{HF}}$ denotes the classical reference energies obtained by \gls{casscf}.
When the superscript is omitted, this implies that no `embedding' method is used, resulting in $E_{\text{CCSD}}$ ($E_{\text{HF}}$) i.e., the \gls{ccsd} (\gls{hf}) energy with all \glspl{mo} as part of the \gls{as}.
Thus, Eq.~\eqref{eq:e_corr} defines a system-independent quality measure allowing us to quantify the quality of the \gls{hf} embedding.
We choose the \gls{ccsd} energy as our `exact' reference because it defines a lower bound on the energy achievable with the \gls{uccsd} Ansatz when all orbitals (and not just a subset) are included in the \gls{as}.

As evident from inspection of Table~\ref{tab:basic_results_overview}, the \gls{hf} embedding leads to significant energy corrections for all of the investigated systems, leading to the recovery of $18-48\%$ of the energy difference between \gls{ccsd} and \gls{hf} (ignoring the case of water in the STO-3G basis where all \glspl{mo} are active).
As discussed previously, the nature of the \gls{as} that gives rise to the largest energy correction cannot be predicted by simple inspection of the \glspl{mo} involved.

\subsection{Density Functional Theory Embedding}
\label{sec:results_DFT_embed}
\gls{ks}-\gls{dft} is the classical effective single-particle method of choice when dealing with molecular systems since both, energies and geometries, are systematically improved over-\gls{hf}.
In order to fully appreciate the advantages of the \gls{dft} embedding scheme, we need to scale up the size of the system beyond what we have done so far with the \gls{hf} approach.
To this end, after a short test on a single water molecule, we introduce a more challenging validation test focusing on the pyridine molecule.

\subsubsection{The water molecule}
\label{sec:results_DFT_embed_water}
As expected, due to the relatively small size of this system, we do not expect important improvements using \gls{dft} over the results obtained with \gls{hf} reported in Section~\ref{sec:results_HF_embed}. 
The same also applies to the shape of the \gls{ks}-\glspl{mo} and the corresponding energy diagram, which do not differ significantly from the HF ones shown in Fig.~\ref{fig:MO_diagram_H2O_631Gst} and~\ref{fig:orbitals_H2O_631Gst}.
\begin{figure}
    \centering
    \includegraphics[width=\columnwidth]{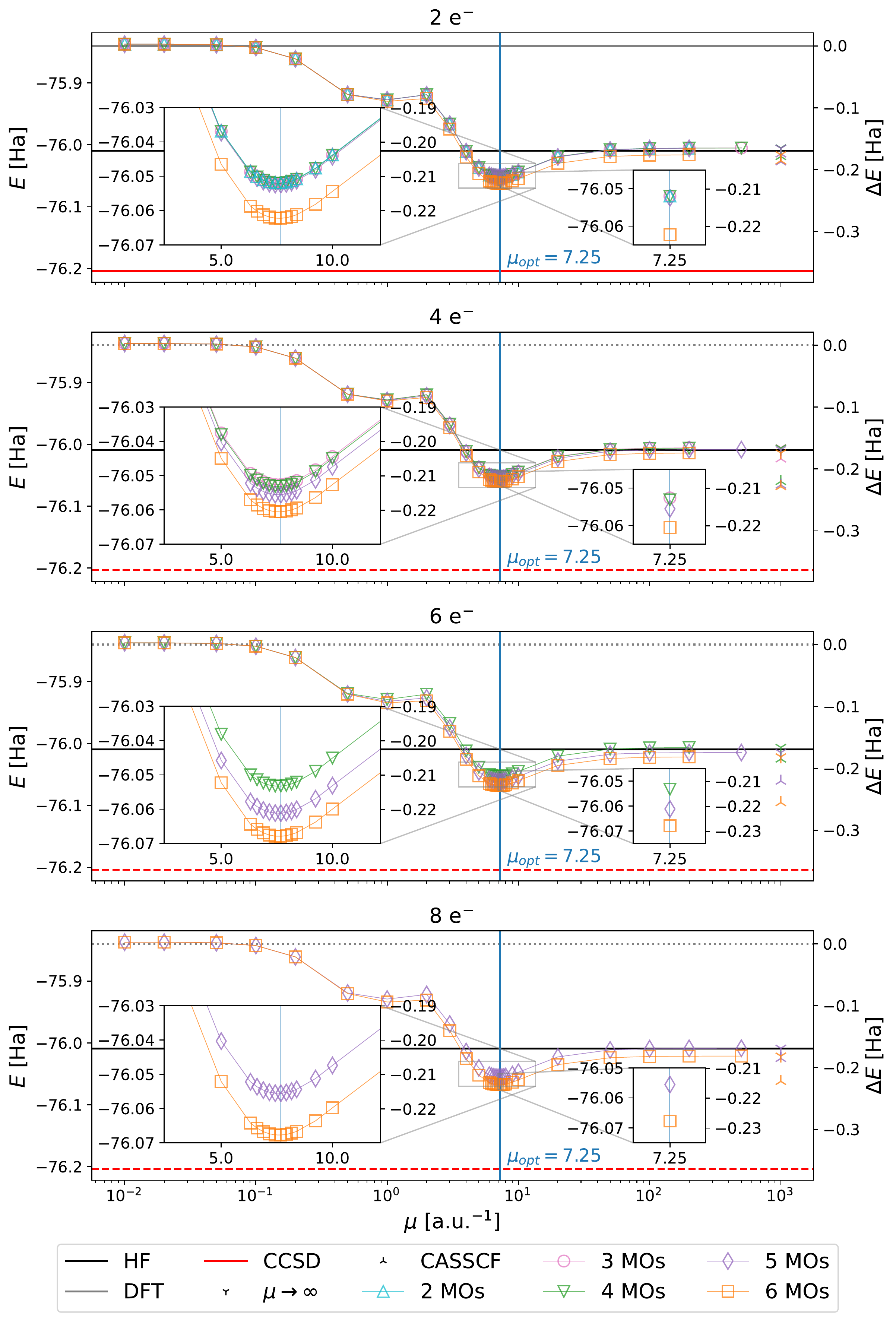}
    \caption{%
    Electronic structure energies (in Hartree) of a \textbf{\ce{H2O} molecule} obtained using the \gls{dft} embedding scheme for a different number of active electrons.
    Each panel groups results obtained for different numbers of active electrons (from the top: 2, 4, 6, 8) while the number of \glspl{mo} in the \gls{as} is color-coded (see legend).
    The classical references were computed using \pyscf{} and the 6-31G* basis.
    The initial density was obtained using \gls{rks}-\gls{dft} with the \gls{lda}/\gls{vwn} \gls{xcf} (grey, dotted).
    For completeness, we also provide the \gls{hf} and \gls{ccsd} references in solid black and dashed red, respectively.
    The range separation parameter, $\mu$, is varied along the $x$ axis.
    Reference values for $\mu\rightarrow\infty$ and for the \gls{casscf} results are included at $\mu=10^3$ with different marker symbols and color-coded according to the respective \glspl{as}.
    }
    \label{fig:iter_H2O_RKS}
\end{figure}
The results obtained with the \gls{dft} embedding are summarized in Fig.~\ref{fig:iter_H2O_RKS} for different choices of the AS. To select the optimal value of the \gls{rs} parameter, $\mu$ (cf.\@ Eq.~\eqref{eq:gpq_rs}), we performed a scan ranging from $0.01$ to $500$ finding an optimal value at $\mu_{opt}=7.25$. 
The number of iterations to reach convergence of the \gls{dft} embedding (see Fig.~\ref{fig:iterative}) varies for the different values of $\mu$. 
In average, 4 iterations were sufficient to  meet the energy threshold of $10^{-10}$ Ha.
We then investigate the behavior of the energy correction in the neighborhood of $\mu_{opt}$ for different sizes of the AS (i.e., number of \glspl{mo}) and number of active electrons. 
The energy curves as a function of the \gls{rs} parameter, $\mu$ (cf.\@ Eq.~\eqref{eq:gpq_rs}), are grouped into panels with constant numbers of active electrons (2, 4, 6 and 8, respectively).
In each panel, the results are shown for different number of active MOs as indicated in the legend.

As expected, for values of $\mu$ tending to zero, we recover the plain \gls{lda}-\gls{dft} result (the dotted, grey line in Fig.~\ref{fig:iter_H2O_RKS}).
Furthermore, in the limit of large $\mu$, the energy converges towards the value obtained with the \gls{hf} embedding scheme discussed in the previous section.
For convenience, the energy values evaluated in this limit ($\mu \rightarrow \infty$), as well as the corresponding \gls{casscf} energies, are reported at $\mu=1000$ in Fig.~\ref{fig:iter_H2O_RKS}.
These data points are indicated with different marker symbols, yet color-coordinated to match the corresponding \glspl{as}.

In order to allow a quantitative comparison of the \gls{dft} embedding results with the ones obtained with the \gls{hf} embedding and the classical reference, \gls{casscf}, we define a new system-independent measure, similar to Eq.~\eqref{eq:e_corr}, which reads
\begin{align}
    \tilde{\varepsilon}^{emb}_{AS} &= \frac{\tilde{E}^{emb}_{AS} - \tilde{E}_{\text{DFT}}}{E_{\text{CCSD}} - \tilde{E}_{\text{DFT}}} [\%] \, ,
    \label{eq:e_corr_dft}
\end{align}
where the $\sim$ indicates energies computed with \gls{ks}-\glspl{mo} (instead of \gls{hf}-\glspl{mo}) and the \gls{dft} reference energy, $\tilde{E}_{\text{DFT}}$, is the value obtained using the \gls{lda}/\gls{vwn} functional (corresponding to the embedding energy in the limit $\mu \rightarrow 0$).
Following the same notation of Eq.~\eqref{eq:e_corr}, the subscript, $AS$, and superscript, $emb$, indicate the method used to treat the \gls{as} and the `embedding' method, respectively.
This leads to the \gls{dft} embedding energies, $\tilde{E}_{\text{q-UCCSD}}^{\text{DFT}}$ (cf.\@ Eq.~\eqref{eq:rs_as_embedding_final}), the \gls{hf} embedding energies based on \gls{ks}-\glspl{mo}, $\tilde{E}_{\text{q-UCCSD}}^{\text{HF}}$ (corresponding to the embedding energy in the limit $\mu \rightarrow \infty$, cf.\@ Eq.~\eqref{eq:as_embedding}), and the classical reference based on \gls{casscf}, $\tilde{E}_{\text{CASSCF}}^{\text{HF}}$. \\
With the help of this measure, we observe (see Table~\ref{tab:dft_results_overview}) a significant improvement of the \gls{dft}  embedding energy correction over the \gls{hf} one by about $13\%$ compared to the common reference \gls{ccsd} value.
However, the success of the method is highly dependent on the value of the \gls{rs} parameter, $\mu$.
Thus, we recommend a sweep over a reasonable range of $\mu$ values when investigating new systems and/or properties thereof as done in this work.

\subsubsection{The pyridine molecule}

\begin{figure}
    \centering
    \includegraphics{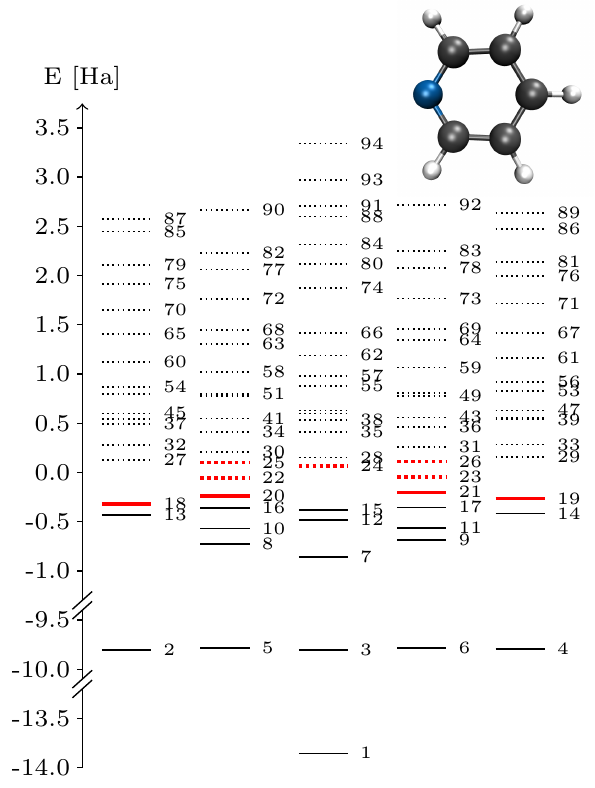}
    \caption{%
    Energy diagram of the \gls{ks}-\glspl{mo} of pyridine in the 6-31G* basis.
    Solid lines correspond to occupied \glspl{mo} while dotted lines represent virtual ones.
    The \glspl{mo} corresponding to the \emph{red} lines ($18$ to $26$) are visualized in Fig.~\ref{fig:orbitals_pyridine_631Gst}.
    To improve readability, every second \gls{mo} label between $40$ and $54$ is omitted.
    }
    \label{fig:MO_diagram_pyridine_631Gst}
\end{figure}
\begin{figure}
    \centering
    \includegraphics[width=\columnwidth]{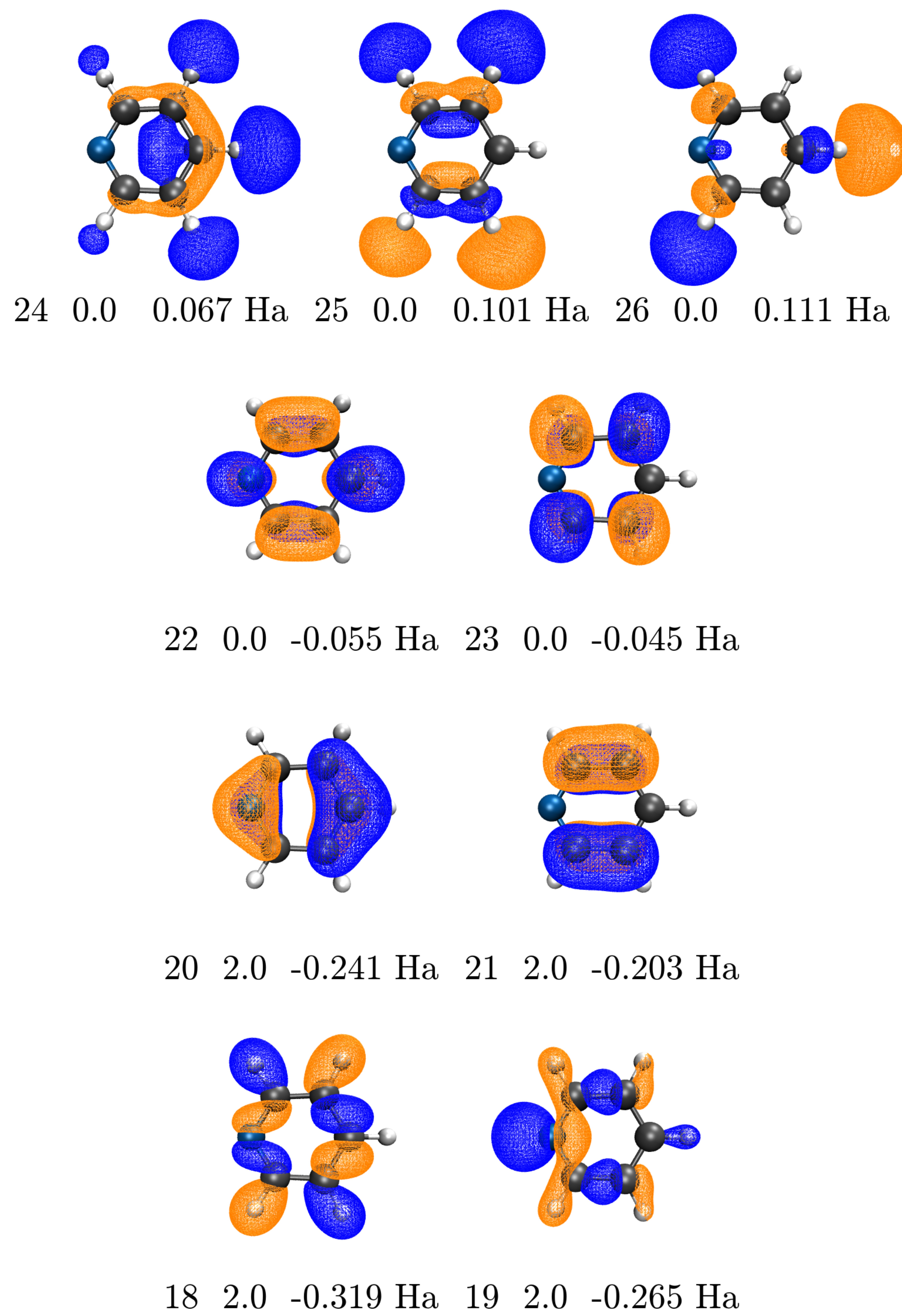}
    \caption{%
    Visualization of the 18th to 26th \gls{ks}-\gls{mo} (out of 94) of pyridine in the 6-31G* basis.
    The orbital coefficients, energies and occupation numbers were obtained using \pyscf{} with a \gls{rks}-\gls{dft} calculation and the \gls{lda}/\gls{vwn} \gls{xcf}.
    All orbitals are rendered using \vmd{}~\cite{Humphrey1996} at an isovalue of $0.05$.
    The three numbers below each orbital correspond to the \gls{hf}-\glspl{mo} index (as indicated in Fig.~\ref{fig:MO_diagram_pyridine_631Gst}), the occupation number, and the orbital energy, respectively.
    }
    \label{fig:orbitals_pyridine_631Gst}
\end{figure}
The pyridine molecule (\ce{C5H5N}) provides a more challenging test case for the validation of the \gls{rs}-\gls{dft} embedding scheme. 
Fig.~\ref{fig:MO_diagram_pyridine_631Gst} presents an energy diagram of the \gls{ks}-\gls{mo} energies of pyridine in the 6-31G* basis and Fig.~\ref{fig:orbitals_pyridine_631Gst} shows the relevant set of orbitals which make up our investigated \glspl{as}.

\begin{figure}
    \centering
    \includegraphics[width=\columnwidth]{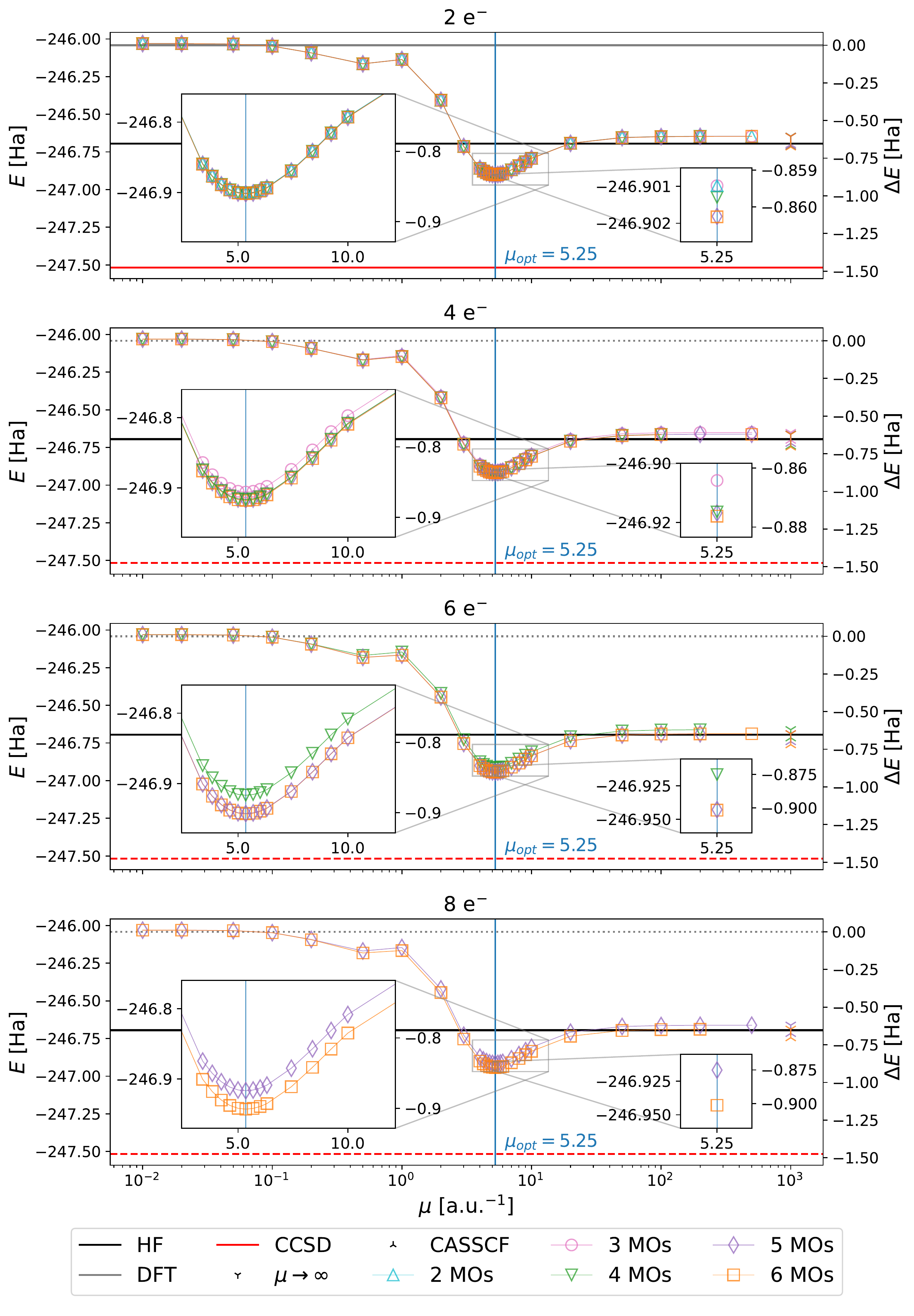}
    \caption{%
    Electronic structure energies (in Hartree) of a \textbf{pyridine} (\ce{C5H5N}) molecule obtained using the \gls{dft} embedding scheme for a different number of active electrons.
    Each panel groups results obtained for different numbers of active electrons (from the top: 2, 4, 6, 8) while the number of \glspl{mo} in the \gls{as} is color-coded (see legend).
    The classical references were computed using \pyscf{} and the 6-31G* basis.
    The initial density was obtained using \gls{rks}-\gls{dft} with the \gls{lda}/\gls{vwn} \gls{xcf} (grey, dotted).
    For completeness, we also provide the \gls{hf} and \gls{ccsd} references in solid black and dashed red, respectively.
    The range separation parameter, $\mu$, is varied along the $x$ axis.
    Reference values for $\mu\rightarrow\infty$ and for the \gls{casscf} results are included at $\mu=10^3$ with different marker symbols and color-coded according to the respective \glspl{as}.
    }
    \label{fig:iter_pyridine_RKS}
\end{figure}
In Fig.~\ref{fig:iter_pyridine_RKS} we summarize the main results obtained for pyridine with the \gls{dft} embedding scheme.
The organization of the four panels follows the same logic used for the case of the water molecule reported in Fig.~\ref{fig:iter_H2O_RKS}: the colors refer to the different sizes of the AS (i.e., number of \glspl{mo} in the \gls{as}) while each panel deals with a different number of active electrons.
The insets in each panel show a zoom of the region around the optimal value, $\mu_{opt}=5.25$.
This differs from the one optimized for water, pointing towards a system dependence of $\mu_{opt}$, which, however, remains constant across different \glspl{as} for the same system.
A similar effect using the \gls{rs}-\gls{lda} functional was already reported in Ref.~\cite{Fromager2007}.
However, the extension to more accurate \glspl{xcf} can alleviate this dependence to some extent~\cite{Fromager2007}.
As for the case of the water molecule, in average 4 iterations between the classical driver (dealing with the DFT environment) and the quantum processor (dealing with the AS) were needed to achieve convergence of the system energy within a threshold of $10^{-10}$~Ha, independently from the value of $\mu$.

Considering the accuracy of the final energies, we observe that the \gls{dft} embedding results outperform the corresponding \gls{hf} ones (at $\mu=1000$) by about $250$mHa, reaching $\tilde{\varepsilon}_{\text{q-UCCSD}}^{\text{DFT}}$ values (cf.\@ Eq.~\eqref{eq:e_corr_dft}) around $60\%$ regardless of the number of active electrons.
However, we have to stress the importance of a parametric sweep of $\mu$ again, since the success of the \gls{dft} embedding strongly depends on the chosen value.
Additionally, a smarter choice of the \gls{as} where the MO are not restricted around the Fermi level may further improve the obtained energy corrections.

\begin{table*}
    \centering
    \caption{%
    Summary of the \gls{dft}-based embedding calculations yielding the largest energy correction, $\tilde{\varepsilon}^{emb}_{AS}$ (cf.\@ Eq.~\eqref{eq:e_corr_dft}). See also Fig.~\ref{fig:iter_H2O_RKS} and~\ref{fig:iter_pyridine_RKS}.
    Reference energies were obtained with the same initial \gls{dft} densities used for the embedding calculation and were computed with \pyscf{}.
    All systems were modeled with the 6-31G* basis. Energy values are in Ha.
    }
    \label{tab:dft_results_overview}
    \begin{tabular}{llrrrrrr}
        \toprule
        \addlinespace%
        System & Active Space & $\mu_{\text{opt}}$ &
        $\tilde{E}_{\text{q-UCCSD}}^{\text{DFT}}$  ($\tilde{\varepsilon}_{\text{q-UCCSD}}^{\text{DFT}}$) &
        $\tilde{E}_{\text{q-UCCSD}}^{\text{HF}}$ ($\tilde{\varepsilon}_{\text{q-UCCSD}}^{\text{HF}}$) &
        $\tilde{E}_{\text{CASSCF}}^{\text{HF}}$ ($\tilde{\varepsilon}_{\text{CASSCF}}^{\text{HF}}$) &
        $\tilde{E}_{\text{DFT}}$ &
        $E_{\text{CCSD}}$ \\
        \addlinespace%
        \midrule
        \ce{H2O} & CAS(2, 6) & $7.25$ & $-76.062$ ($61.0$) & $-76.016$ ($48.4$) & $-76.026$ ($51.1$) & $-75.840$ & $-76.204$ \\
        \ce{H2O} & CAS(4, 6) & $7.25$ & $-76.061$ ($60.7$) & $-76.014$ ($47.8$) & $-76.069$ ($62.9$) & $-75.840$ & $-76.204$ \\
        \ce{H2O} & CAS(6, 6) & $7.25$ & $-76.068$ ($62.6$) & $-76.021$ ($49.7$) & $-76.095$ ($70.1$) & $-75.840$ & $-76.204$ \\
        \ce{H2O} & CAS(8, 6) & $7.25$ & $-76.068$ ($62.6$) & $-76.021$ ($49.7$) & $-76.062$ ($61.0$) & $-75.840$ & $-76.204$ \\
        pyridine & CAS(2, 6) & $5.25$ & $-246.902$ ($58.3$) & $-246.647$ ($41.0$) & $-246.717$ ($45.8$) & $-246.042$ & $-247.517$ \\
        pyridine & CAS(4, 6) & $5.25$ & $-246.918$ ($59.4$) & $-246.663$ ($42.1$) & $-246.747$ ($47.8$) & $-246.042$ & $-247.517$ \\
        pyridine & CAS(6, 6) & $5.25$ & $-246.943$ ($61.1$) & $-246.689$ ($43.9$) & $-246.761$ ($48.7$) & $-246.042$ & $-247.517$ \\
        pyridine & CAS(8, 6) & $5.25$ & $-246.943$ ($61.1$) & $-246.688$ ($43.8$) & $-246.745$ ($47.7$) & $-246.042$ & $-247.517$ \\
        \bottomrule
    \end{tabular}
\end{table*}

An encouraging observation that we can draw from inspection of Fig.~\ref{fig:iter_pyridine_RKS} is that for a system of the size of pyridine (with $42$ electrons) we can already outperform the classical \gls{casscf} approach for all \gls{as} choices.
In fact, we obtain energy corrections of around $60\%$ compared to values around $48\%$.
This is a very promising result, which highlights the benefits that can be obtained from the combination of \gls{rs} embedding approaches with quantum electronic structure algorithms.
We summarize these results in Table~\ref{tab:dft_results_overview}, where we compare the energy corrections $\tilde{\varepsilon}^{emb}_{AS}$ (cf.\@ Eq.~\eqref{eq:e_corr_dft}) obtained with the \gls{dft} and \gls{hf} embedding schemes as well as the classical reference, \gls{casscf}.

It should be noted that the \gls{dft} embedding scheme requires multiple \gls{vqe} calculations due to its iterative nature.
While this poses a computational burden, the potential benefits in terms of accuracy that can be obtained for more complex systems will be important and will justify the additional computational costs.

\section{Conclusions}
\label{sec:conclusions}
In this work, we introduced an embedding scheme that enables the partitioning of electronic structure calculations into an \glsfirst{as} subsystem treated with a high level quantum algorithm and an \textit{environment} described at the \gls{hf} or \gls{dft} level of theory.
In this way, we can restrict the quantum calculations to a critical subset of molecular orbitals that can fit on state-of-the-art quantum computers, while the remaining electrons provide the embedding potential computed using a classical algorithm. 
Since in most chemical processes, the quality of the electronic structure predictions depends on a small set of frontier orbitals, this scheme will allow the solution of interesting quantum chemistry problems where the \gls{as} can be described with a quantum algorithm presenting a favorable scaling in the number of active electrons.
We show the performance of the embedding scheme in the case of a few test molecular systems, namely \ce{H2O}, \ce{N2}, \ce{CH2}, \ce{O2} and pyridine, highlighting the benefits of the recursive update of the embedding potential for an improved convergence of the computed ground state energies.
It is important to mention that the use of the iterative Range-Separated \gls{dft} embedding requires the tuning of an extra parameter, which cannot be set \textit{a priori}.
Further investigation is needed to automatize this technique for general use in larger molecular systems.  

Of particular relevance are the results obtained for pyridine;
in this case we showed that the iterative quantum computing embedding scheme is able to outperform classical active space methods such as \gls{casscf} using a reasonable small number of quantum resources 
(i.e., the \gls{dft} embedding scheme with the \gls{uccsd} expansion of the AS recovers $13\%$ more energy than the \gls{casscf} approach with the same \gls{as}).

Improvements on the proposed embedding scheme can be obtained through the combination of the iterative update of the embedding potential together with the simultaneous optimization of the active orbitals as done, for instance, in the multiconfigurational self-consistent field (MCSCF) approach. 

We believe that the proposed \gls{hf} and \gls{dft} embedding schemes will provide a fundamental framework for the scaling up of quantum electronic structure calculations to large molecular systems with an arbitrary number of electrons (i.e., as many as a \gls{hf} or \gls{dft} calculations can deal with). 
The possibility of partitioning the solution of the electronic structure problem into an \textit{active} component (defined by the AS) treated by means of a quantum computing algorithm and an \textit{inert} \textit{environment} component solved at the \gls{hf} or \gls{dft} level of theory will open up new avenues for the use of quantum computers in the solution of important problems in physics, chemistry, biology and medicine.
 

\section*{Acknowledgements}
The authors thank Valery Weber and J{\"u}rg Hutter for useful discussions as well as Manfred Sigrist who advised M. R. during a significant part of this work. \\
I.T. and P.J.O acknowledge  financial  support  from  the Swiss  National  Science  Foundation  (SNF)  through grant No.  200021-179312.

\section*{AIP Publishing Data Sharing Policy}
The data that support the findings of this study are available from the corresponding author upon reasonable request.

\appendix

\section{Hartree--Fock Embedding}
\label{app:hf_embed_deriv}

In this section we provide more detailed derivations of the \gls{hf} embedding.
First, we derive the restricted spin case and generalize the equations for unrestricted spins in the second part of this section.

\subsection{Restricted Spins}
\label{app:rhf_embed_deriv}

We introduce the splitting of the one--electron density matrix, $D=D^{A}+D^{I}$, into Eq.~\eqref{eq:schroedinger_2nd_quant} one term at a time.
The simplest case is the one of the one--electron contribution which becomes
\begin{align}
    \sum_{pq} h_{pq}D_{pq}
      &= \sum_{vq} h_{vq}D_{vq}^{A} + \sum_{jq} h_{jq}D_{jq}^{I} \nonumber \\
      &= \sum_{uv} h_{uv}D_{uv}^{A} + 2 \sum_{j} h_{jj}, \label{eq:1electron_terms_si}
\end{align}
where we use the fact that a density matrix element vanishes when any of its indices correspond to a virtual orbital.
We can proceed analogously with the two-electron terms as in
\begin{align}
    \frac{1}{2} \sum_{pqrs} g_{pqrs}d_{pqrs}
      &= \frac{1}{2} \sum_{pqjs} g_{pqjs}d_{pqjs} + \frac{1}{2} \sum_{jqus} g_{jqus}d_{jqus} \nonumber \\
      &+ \frac{1}{2} \sum_{vqus} g_{vqus}d_{vqus}. \label{eq:2electron_terms_full_si}
\end{align}
We can now express the two--electron density matrices in terms of one--electron ones,
\begin{subequations}
    \begin{align}
        d_{pqjs} &= D_{pq}D_{js} - \delta_{qj}D_{ps} \nonumber \\
                 &= 2\delta_{js}D_{pq} - \delta_{qj}\delta_{sq}D_{pq} \nonumber \\
                 &= (2\delta_{js} - \delta_{qj}\delta_{sq}) D_{pq} \\
        \shortintertext{and}
        d_{jqus} &= d_{usjq} = D_{us}D_{jq} - \delta_{sj}D_{uq} \nonumber \\
                 &= 2\delta_{jq}D_{us} - \delta_{sj}D_{uq} \nonumber \\
                 &= 2\delta_{qj}\delta_{sq}D_{uq} - \delta_{sj}D_{uq} \nonumber \\
                 &= (2\delta_{qj}\delta_{sq} - \delta_{sj}) D_{uq},
    \end{align}
    \label{eq:2rdm_to_1rdm_si}%
\end{subequations}
where we omit the superscripts $I$ and $A$ for brevity.
These expressions can then be inserted into Eq.~\eqref{eq:2electron_terms_full_si} to obtain
\begin{align}
    \frac{1}{2} \sum_{pqrs} g_{pqrs}d_{pqrs}
      &= \sum_{pqj} g_{pqjj}D_{pq} - \frac{1}{2} \sum_{pqj} g_{pjjq}D_{pq} \nonumber \\
      &+ \sum_{jqu} g_{jjuq}D_{uq}^{A} - \frac{1}{2} \sum_{jqu} g_{jquj}D_{uq}^{A} \nonumber \\
      &+ \frac{1}{2} \sum_{uvxy} g_{uvxy}d_{uvxy}^{A}. \label{eq:2electron_terms_si}
\end{align}
Upon inspection, it becomes clear that the first two sums only yield non-zero contributions when $p=q=k$ or when $p=u$ and $q=v$.
In the latter case, this causes the sums to coincide with the third and forth term.
These observations allow us to simplify Eq.~\eqref{eq:2electron_terms_si} to become
\begin{align}
    \frac{1}{2} \sum_{pqrs} g_{pqrs}d_{pqrs}
      &= 2 \sum_{kj} (g_{kkjj} - \frac{1}{2} g_{kjjk}) \nonumber \\
      &+ 2 \sum_{jvu} (g_{jjuv} - \frac{1}{2} g_{jvuj}) D_{uv}^{A} \nonumber \\
      &+ \frac{1}{2} \sum_{uvxy} g_{uvxy}d_{uvxy}^{A}. \label{eq:2electron_terms_simplified_si}
\end{align}
Finally, we can substitute Eq.~\eqref{eq:1electron_terms_si} and Eq.~\eqref{eq:2electron_terms_simplified_si} into Eq.~\eqref{eq:schroedinger_2nd_quant} yielding
\begin{align}
    E &= \sum_{j} \left[ 2h_{jj} + \sum_{k} (2g_{kkjj} - g_{kjjk}) \right] \nonumber \\
      &+ \sum_{uv} \left[ h_{uv} + \sum_{j} (2g_{jjuv} - g_{jvuj}) \right] D_{uv}^{A} \nonumber \\
      &+ \frac{1}{2} \sum_{uvxy} g_{uvxy}d_{uvxy}^{A}. \label{eq:as_embedding_full_si}
\end{align}
This equation simplifies to Eq.~\eqref{eq:as_embedding} with the use of the \emph{inactive Fock operator}, Eq.~\eqref{eq:fock_inactive}, and the \emph{inactive energy}, Eq.~\eqref{eq:energy_inactive}.

\subsection{Unrestricted Spins}
\label{app:uhf_embed_deriv}

For unrestricted spins we have to remove the implicit summation over the spin state, $\sigma$, and calculate with the one-- and two--electron density matrices for each spin separately,
\begin{subequations}
    \begin{align}
        D_{pq}^{\sigma} &= \Braket{\Psi|\hat{E}_{pq}^{\sigma}|\Psi}
        &\forall\sigma\in\{\alpha,\beta\} \\
        d_{pqrs}^{\sigma\tau} &= \Braket{\Psi|\hat{e}_{pqrs}^{\sigma\tau}|\Psi}
        &\forall\sigma,\tau\in\{\alpha,\beta\} \\
        \hat{E}_{pq}^{\sigma} &= \crop{p\sigma}\anop{q\sigma}
        &\forall\sigma\in\{\alpha,\beta\} \\
        \hat{e}_{pqrs}^{\sigma\tau} &= \left\{
        \begin{array}{lrl}
             \hat{E}_{pq}^{\sigma}\hat{E}_{rs}^{\sigma}-\delta_{qr}\hat{E}_{ps}^{\sigma}
             &, &\sigma=\tau \\
             \hat{E}_{pq}^{\sigma}\hat{E}_{rs}^{\tau} &, &\sigma\neq\tau
        \end{array}
        \right.&\forall\sigma,\tau\in\{\alpha,\beta\}
        \label{eq:2el_exc_op_us_si}
    \end{align}
\end{subequations}
Thus, in contrast to the closed shell description with restricted spins in the previous section, we now have to keep track of the spin state which each index iterates over.
This is indicated by the additional superscripts, $\sigma$ and $\tau$, and summation labels, $n_\sigma$ and $n_\tau$, respectively.
For brevity we refrain from explicitly denoting that $\sigma,\tau\in\{\alpha,\beta\}$ for the remainder of this derivation.

Analogous to Eq.~\eqref{eq:1electron_terms_si} the one--electron contribution can be written per spin state as
\begin{align}
    \sum_{pq}^{n_\sigma}h_{pq}D_{pq}^{\sigma}
    &= \sum_{uv}^{n_\sigma} h_{uv}D_{uv}^{A,\sigma}
    + \sum_{j}^{n_\sigma} h_{jj}.
    \label{eq:1electron_terms_us_si}
\end{align}
For the two--electron contributions we have to differentiate between two cases.
In the first case, the spins of both electrons are aligned, $\sigma=\tau$, and the resulting equation can be derived in full analogy to Eq.~\eqref{eq:2electron_terms_simplified_si},
\begin{align}
    \frac{1}{2} \sum_{pq}^{n_\sigma}\sum_{rs}^{n_\sigma} g_{pqrs}d_{pqrs}^{\sigma\sigma}
      &= \frac{1}{2} \sum_{k}^{n_\sigma}\sum_{j}^{n_\sigma} (g_{kkjj} - g_{kjjk}) \nonumber \\
      &+ \sum_{j}^{n_\sigma}\sum_{uv}^{n_\sigma} (g_{jjuv} - g_{jvuj}) D_{uv}^{A,\sigma} \nonumber \\
      &+ \frac{1}{2} \sum_{uv}^{n_\sigma}\sum_{xy}^{n_\sigma} g_{uvxy}d_{uvxy}^{A,\sigma\sigma}.
      \label{eq:2electron_terms_us_aligned_si}
\end{align}
The second case of opposite spins, $\sigma\neq\tau$, behaves slightly different due to the differing two--electron excitation operator, Eq.~\eqref{eq:2el_exc_op_us_si}.
Thus, the expression of the two--electron density matrices in terms of one--electron ones analogous to Eq.~\eqref{eq:2rdm_to_1rdm_si} becomes
\begin{subequations}
    \begin{align}
        d_{pqjs}^{\sigma\tau} &= D_{pq}^{\sigma}D_{js}^{\tau} = \delta_{js}D_{pq}^{\sigma}, \\
        \shortintertext{and}
        d_{jqus}^{\sigma\tau} &= D_{jq}^{\sigma}D_{us}^{\tau} = \delta_{jq}D_{us}^{\tau},
    \end{align}
\end{subequations}
which leads to the first expression of the two--electron contribution of opposite spins
\begin{align}
    \frac{1}{2} \sum_{pq}^{n_\sigma}\sum_{rs}^{n_\tau} g_{pqrs}d_{pqrs}^{\sigma\tau}
    &= \frac{1}{2} \sum_{pq}^{n_\sigma}\sum_{j}^{n_\tau}
    g_{pqjj}D_{pq}^{\sigma} \nonumber \\
    &+ \frac{1}{2} \sum_{j}^{n_\sigma}\sum_{us}^{n_\tau}
    g_{jjus}D_{us}^{A,\tau} \nonumber \\
    &+ \frac{1}{2} \sum_{vq}^{n_\sigma}\sum_{us}^{n_\tau}
    g_{vqus}d_{vqus}^{A,\sigma\tau}.
\end{align}
Using the same arguments as before we can once again simplify this equation to become
\begin{align}
    \frac{1}{2} \sum_{pq}^{n_\sigma}\sum_{rs}^{n_\tau}~&g_{pqrs}d_{pqrs}^{\sigma\tau} = \nonumber \\
    &= \frac{1}{2} \sum_{k}^{n_\sigma}\sum_{j}^{n_\tau}
    g_{kkjj}
    + \frac{1}{2} \sum_{uv}^{n_\sigma}\sum_{j}^{n_\tau}
    g_{uvjj}D_{uv}^{A,\sigma} \nonumber \\
    &+ \frac{1}{2} \sum_{j}^{n_\sigma}\sum_{uv}^{n_\tau}
    g_{jjuv}D_{uv}^{A,\tau}
    + \frac{1}{2} \sum_{uv}^{n_\sigma}\sum_{xy}^{n_\tau}
    g_{uvxy}d_{uvxy}^{A,\sigma\tau}.
    \label{eq:2electron_terms_us_opposite_si}
\end{align}
Finally, we can obtain the two--electron contributions for unrestricted spins by combining Eq.~\eqref{eq:2electron_terms_us_aligned_si} and Eq.~\eqref{eq:2electron_terms_us_opposite_si} into
\begingroup
\allowdisplaybreaks
\begin{align}
    \frac{1}{2} \sum_{pqrs}~&g_{pqrs}d_{pqrs} = \nonumber \\
    &= \frac{1}{2} \sum_{j}^{n_\alpha} \left[
        \sum_{k}^{n_\alpha} (g_{kkjj} - g_{kjjk})
        + \sum_{k}^{n_\beta} g_{kkjj}
    \right] \nonumber \\
    &+ \frac{1}{2} \sum_{j}^{n_\beta} \left[
        \sum_{k}^{n_\beta} (g_{kkjj} - g_{kjjk})
        + \sum_{k}^{n_\alpha} g_{kkjj}
    \right] \nonumber \\
    &+ \sum_{uv}^{n_\alpha} \left[
        \sum_{j}^{n_\alpha} (g_{uvjj} - g_{ujjv})
        + \sum_{j}^{n_\beta} g_{uvjj}
    \right] D_{uv}^{A,\alpha} \nonumber \\
    &+ \sum_{uv}^{n_\beta} \left[
        \sum_{j}^{n_\beta} (g_{uvjj} - g_{ujjv})
        + \sum_{j}^{n_\alpha} g_{uvjj}
    \right] D_{uv}^{A,\beta} \nonumber \\
    &+ \frac{1}{2} \sum_{uvxy} g_{uvxy} d_{uvxy}^{A}.
\end{align}
\endgroup

In full analogy to the restricted spin case this allows us to define the \emph{inactive} Fock operator and energy as
\begin{align}
    F^{I,\sigma}_{pq} &= h_{pq}
    + \sum_{i}^{n_\sigma} (g_{iipq} - g_{iqpi})
    + \sum_{i}^{n_\tau} g_{iipq} \\
    \shortintertext{and}
    E^{I} &= \frac{1}{2} \sum_{\sigma=\alpha}^{\beta} \sum_{j} h_{jj} + F_{jj}^{I,\sigma} \nonumber \\
    &= \frac{1}{2} \sum_{\sigma=\alpha}^{\beta} \sum_{ij} (h_{ij} + F_{ij}^{I,\sigma}) D_{ij}^{I,\sigma}.
\end{align}

\section{Density Functional Theory Embedding}
\label{app:dft_embed_deriv}

In this section, we provide more detailed steps deriving the embedding equations of the iterative \gls{dft} embedding scheme.
In doing so, we follow the work of Hedeg{\aa}rd et.\@ al.\@ \cite{Hedegard2015} rather closely.
We focus on the steps necessary to arrive at the final form of the total electronic energy, Eq.~\eqref{eq:rs_as_embedding_final}, after the introduction of the linear model, Eq.~\eqref{eq:lin_mod_approximation}.

We start by noting that $\Delta D_{pq}^{(i)} = \Delta D^{A,(i)}_{uv}$ since the inactive part of the density matrix, $D^{I}_{ij}$, is constant by definition.
Thus, we can express the total electronic energy after exploiting inherent properties of the one--electron density matrices as
\begin{align}
    E &= E^{I,\text{LR}} + E^{A,(i+1),\text{LR}} + E_{\text{coul}+\text{xc}}^{\text{SR}}\left[\rho^{(i)}\right] \nonumber \\
    &+ \sum_{uv} \left(j^{(i),\text{SR}}_{uv} + \nu^{\text{SR}}_{\text{xc},uv}\left[\rho^{(i)}\right]\right) \Delta D^{A,(i)}_{uv}.
    \label{eq:rs_as_embedding_si}
\end{align}
To ease the implementation of Eq.~\eqref{eq:rs_as_embedding_si} we can rewrite the equation and group its terms into active and inactive ones.
To do so, we start by rewriting the Coulomb part of the third term by expanding $D^{(i)} = D^{I} + D^{A,(i)}$ twice,
\begin{align}
    J^{\text{SR}}\left[\rho^{(i)}\right] &= \frac{1}{2} \sum_{pqrs} D^{(i)}_{pq} g^{\text{SR}}_{pqrs} D^{(i)}_{rs} \nonumber \\
        &= \frac{1}{2} \sum_{pq} D^{(i)}_{pq} \Big(
            \sum_{ij} g^{\text{SR}}_{pqij} D^{I}_{ij} + \sum_{uv} g^{\text{SR}}_{pquv} D^{A,(i)}_{uv}
        \Big) \nonumber \\
        &= \frac{1}{2} \Big(
            \sum_{klij} D^{I}_{kl} g^{\text{SR}}_{klij} D^{I}_{ij} +
            \sum_{xyij} D^{A,(i)}_{xy} g^{\text{SR}}_{xyij} D^{I}_{ij} \nonumber \\
        &+ \sum_{kluv} D^{I}_{kl} g^{\text{SR}}_{kluv} D^{A,(i)}_{uv} +
            \sum_{xyuv} D^{A,(i)}_{xy} g^{\text{SR}}_{xyuv} D^{A,(i)}_{uv} \Big).
            \label{eq:rs_as_expand_coulomb_third_si}
\end{align}
We can proceed analogously with the Coulomb part of the fourth term,
\begin{align}
    \sum_{uv} &j^{(i),\text{SR}}_{uv} \Delta D^{A,(i)}_{uv}
    = \sum_{pquv} D^{(i)}_{pq} g^{\text{SR}}_{pquv} \left( D^{A,(i+1)}_{uv} - D^{A,(i)}_{uv} \right) \nonumber \\
    &= \sum_{ijuv} D^{I}_{ij} g^{\text{SR}}_{ijuv} D^{A,(i+1)}_{uv} - D^{I}_{ij} g^{\text{SR}}_{ijuv} D^{A,(i)}_{uv} \nonumber \\
    &+ \sum_{xyuv} D^{A,(i)}_{xy} g^{\text{SR}}_{xyuv} D^{A,(i+1)}_{uv} - D^{A,(i)}_{xy} g^{\text{SR}}_{xyuv} D^{A,(i)}_{uv}.
    \label{eq:rs_as_expand_coulomb_fourth_si}
\end{align}
By gathering and canceling matching terms of Eq.~\eqref{eq:rs_as_expand_coulomb_third_si} and Eq.~\eqref{eq:rs_as_expand_coulomb_fourth_si} through the use of the symmetry, $g_{pqrs} = g_{rspq}$, we arrive at the final expression of the \gls{sr} Coulomb contributions
\begin{align}
    J^{\text{SR}}&\left[\rho^{(i)}\right] + \sum_{uv} j^{(i),\text{SR}}_{uv} \Delta D^{A,(i)}_{uv} \nonumber \\
    &= \frac{1}{2} \sum_{ij} j^{I,\text{SR}}_{ij} D^{I}_{ij} - \frac{1}{2} \sum_{uv} j^{A,(i),\text{SR}}_{uv} D^{A,(i)}_{uv} \nonumber \\
    &+ \sum_{uv} j^{I,\text{SR}}_{uv} D^{A,(i+1)}_{uv} + \sum_{uv} j^{A,(i),\text{SR}}_{uv} D^{A,(i+1)}_{uv}.
    \label{eq:rs_as_expand_coulomb_final_si}
\end{align}
Inserting Eq.~\eqref{eq:rs_as_expand_coulomb_final_si} into Eq.~\eqref{eq:rs_as_embedding_si} finally leads to
\begin{align}
    E &= E^{I,\text{LR}} + E_{\text{xc}}^{\text{SR}}\left[\rho^{(i)}\right]
    + \frac{1}{2} \sum_{ij} j^{I,\text{SR}}_{ij} D^{I}_{ij} \nonumber \\
    &- \frac{1}{2} \sum_{uv} j^{A,(i),\text{SR}}_{uv} D^{A,(i)}_{uv}
    - \sum_{uv} \nu^{\text{SR}}_{\text{xc},uv}\left[\rho^{(i)}\right] D^{A,(i)}_{uv}
    \nonumber \\
    &+ E^{A,\text{LR}}
    + \sum_{uv} j^{I,\text{SR}}_{uv} D^{A,(i+1)}_{uv}
    + \sum_{uv} j^{A,(i),\text{SR}}_{uv} D^{A,(i+1)}_{uv} \nonumber \\
    &+ \sum_{uv} \nu^{\text{SR}}_{\text{xc},uv}\left[\rho^{(i)}\right] D^{A,(i+1)}_{uv},
    \label{eq:rs_as_embedding_split_si}
\end{align}
where we have re-ordered the terms such that the upper two lines contain all the inactive terms and the lower lines all the active ones.
After insertion of the expressions for the \gls{lr} energy contributions, $E^{I}$ and $E^{A}$, according to Eq.~\eqref{eq:energy_inactive} and Eq.~\eqref{eq:as_embedding}, respectively, we arrive at the final expression of the total electronic energy, Eq.~\eqref{eq:rs_as_embedding_final}.

The extension of these equations to unrestricted spins is similarly straight forward as in the case of the \gls{hf} embedding (cf.\@ Appendix~\ref{app:uhf_embed_deriv}).

\bibliographystyle{jcp}

\end{document}